# Accelerating the development of oxynitride thin films: A combinatorial investigation of the Al-Si-O-N system

Stefanie Frick[1], Oleksandr Pshyk[1], Arnold Müller[2], Alexander Wieczorek[1], Kerstin Thorwarth[1], Sebastian Siol[1*]

[1] Empa, Swiss Federal Laboratories for Materials Science and Technology, Dübendorf, Switzerland

[2] Laboratory of Ion Beam Physics, ETH Zurich, Zürich, Switzerland

*Corresponding author:*
*Sebastian Siol, Sebastian.Siol@empa.ch*

**Abstract**

Oxynitrides are used in a variety of applications including photocatalysts, high-k dielectrics or wear-resistant coatings and often show intriguing multi-functionality. To accelerate the co-optimization of the relevant material properties of these compositionally complex oxynitride systems, high-throughput synthesis and characterization methods are desirable. In the present work, three approaches were investigated to obtain orthogonal anion and cation gradients on the same substrate by magnetron sputtering. The different approaches included varying positions of the local reactive gas inlets and different combinations of target materials. The best performing approach was applied to screen a large two-dimensional area of the quaternary phase space within the Al-Si-O-N system. This material system is a promising candidate for transparent protective coatings with variable refractive indices. With only five depositions of combinatorial libraries, an anion composition range of 2-46% O/(N+O) and a cation composition range of 4-44% Si/(Al+Si) is covered. For lower oxygen and silicon contents, a region with hardness of up to 25 GPa is observed, where the material exhibits either wurtzite AlN or a composite microstructure. By increasing the deposition temperature to 400 °C, an extension of this region can be achieved. At higher oxygen and silicon contents, the structure of the samples is X-ray amorphous. In this structural region, an intimate correlation between hardness and refractive index is confirmed. The results of this study introduce a practical approach to perform high-throughput development of mixed anion materials, which is transferable to many materials systems and applications.

**Graphical abstract**

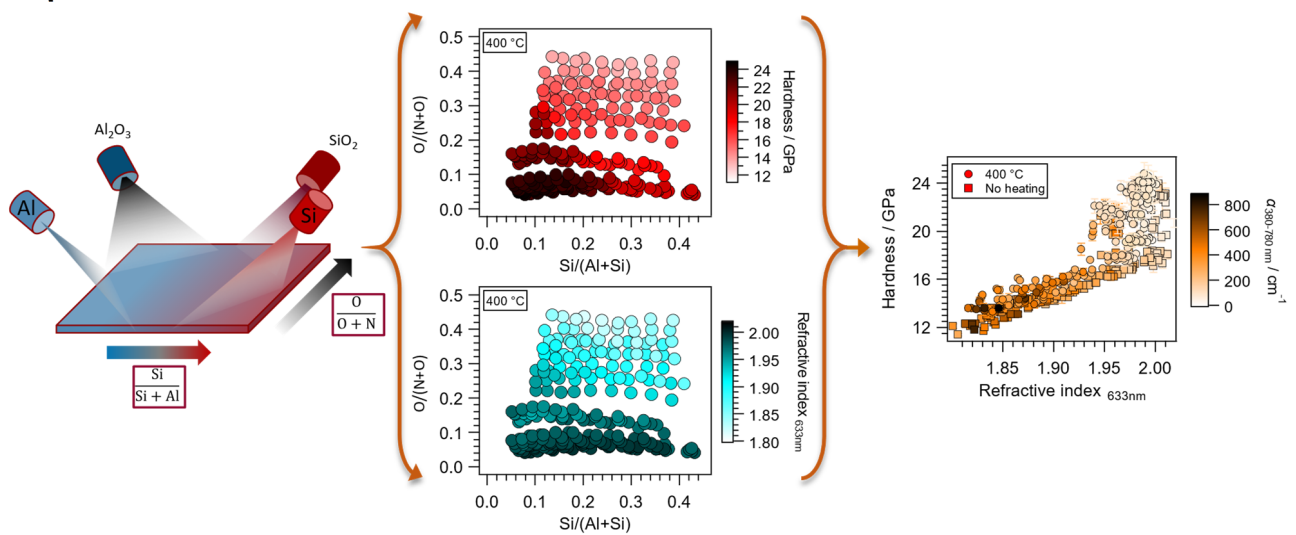

# 1. Introduction

Oxynitrides are a versatile class of materials with a broad range of applications, ranging from high k-dielectrics [1–3] over photocatalysts [4], pigments [5,6], and phosphors [7] to solar-selective absorbers [8], anti-reflection [9] and protective coatings [10,11] or gas diffusion barriers [3,12]. This versatility is enabled by the tunability of the properties of oxynitrides via the anion composition. Properties like electrical conductivity [13], band gap [5], optical constants [8,11], and mechanical properties like hardness and elastic constants [11,13,14] or biomedical compatibility [15,16] can vary over a significant range from the oxide's to the nitride's characteristics. A prominent example is the transition from the opaque metallic TiN to transparent $TiO_2$ with a band gap of >3 eV [17,18].

The compositional complexity particularly in quaternary oxynitrides *AB*ON (*A,B* indicating metal ions) makes their development via trial-and-error inherently time-consuming. Combinatorial workflows using gradient deposition promise rapid materials development [19]. However, today, this type of composition screening is typically limited to gradients in cation composition, even in the case of oxynitride coatings [20–25]. The ability to perform a two-dimensional screening by simultaneously obtaining cation (*A-B*) and anion (O-N) gradients can accelerate the development of oxynitrides significantly. Reports about the combinatorial screening of anions are generally rare and often limited to heavier anions, e.g. Se and Te [26,27]. Only a few groups have reported on intentional gradients in either the nitrogen or the oxygen content. These were achieved in magnetron sputtering via an asymmetric $N_2$ gas inlet [28], varying $O_2$ flows with time [29], or differently oxidized precursors in molecular beam epitaxy (MBE) [30]. Gradients in the oxygen-to-nitrogen ratio, as of interest for oxynitrides, have been reported for chemical vapor deposition (CVD) [31,32] and pulsed laser deposition (PLD) [33] combinatorial depositions, while keeping the cation stoichiometry fixed. To the best of our knowledge, combinatorial gradients of the oxygen-to-nitrogen ratio in magnetron sputtering has been only reported for co-sputtering of oxide and nitride compounds with different cations, which lead to overlaying cation and anion gradients [34], [35], and consequently does not allow for a decoupling of the effects of cation and anion composition. Instead, it would be of high interest to investigate approaches that enable to obtain controllable O/N-gradients perpendicular to a cation gradient to analyze the dependencies on cation and anion stoichiometry independently and to maximize the area covered in quaternary oxynitride phase spaces.

For a proof of principle, the quaternary Al-Si-O-N system is a promising candidate, since the ternary systems Al-Si-N [36–39] and Al-O-N [11] are already well investigated for magnetron sputtered coatings. Moreover, this material system is promising for the application as a multi-functional protective anti-reflection coating, where mechanically robust, low refractive index materials are of interest. The ternary systems are reported to exhibit compositional regimes with a nanocomposite microstructure [11,38,40] with a hardness of up to 32 GPa in the Al-Si-N system [37]. Additionally, the refractive index can be tuned between 1.6 and 2.2 via the oxygen-to-nitrogen ratio in the Al-O-N system [11,41]. The current limitation in the applicability of Al-O-N for durable anti-reflection coatings is the reported decrease in hardness accompanying a reduction of the refractive index [11,41]. To investigate whether an independent control of hardness and refractive index is possible in the Al-Si-O-N system, a comprehensive 2D screening of the phase space would be necessary.

In this work, we compare different methods to obtain orthogonal anion and cation gradients and apply them to screen the phase space of the quaternary Al-Si-O-N system. For this purpose, we investigated three different setups to obtain O/N-gradients via reactive magnetron sputtering, including local gas inlets and varying target material combinations. We find that the combination of metallic and oxide targets sputtered in an $N_2$ atmosphere allows for the best control in obtaining independent anion and cation gradients. Applying this approach, we screen anion compositions of 2 – 46% O/(N+O) and cation compositions of 4 – 44% Si/(Al+Si) and study the impact of anion and cation composition on hardness and refractive index. In addition, we investigate the influence of varying deposition temperatures on the optical and mechanical properties of the coatings. We find an extension of the high hardness "composite" phase towards higher silicon and oxygen contents with increasing deposition temperature. Outside this region, our results confirm an intimate coupling between refractive index and hardness as documented in prior studies. The results of our work introduce a practical approach to perform combinatorial screening of compositionally complex mixed-anion systems, which is broadly applicable.

## 2. Materials and methods

### 2.1. Thin film deposition

Thin film material libraries were deposited by reactive magnetron co-sputtering from two to four targets in an AJA ATC-1500 sputter tool with confocal sputter-up geometry. In **Fig. 1**, the target configurations and locations of the reactive gas inlets for different strategies for achieving compositional gradients are depicted. A detailed description of the different approaches is provided in the Results and Discussion section. The following 2" targets were used depending on the deposition approach: Aluminum (99.9995% purity, Kurt J. Lesker Company Ltd.) and B-doped silicon (99.999% purity, HMW Hauner GmbH & Co. KG as well as an alloy target Al/Si 80 at.%/20 at.% (Plansee) and ceramic $Al_2O_3$ and $SiO_2$ targets (99.99% purity, Kurt J. Lesker Company Ltd.). High power impulse magnetron sputtering (HiPIMS) was applied to the metallic Al target to support densification of the coatings, while direct current magnetron sputtering (DCMS) was used for the metallic Si target. Applying this hybrid Al-HiPIMS/Si-DCMS approach, Lewin *et al.* have reported an increase in hardness compared to coatings deposited with Al-DCMS/Si-DCMS [42], which was observed in preliminary studies of our group as well. The DC power applied to the metallic Si target was 15-25 W, while the HiPIMS power of 50-100 W was applied to the Al target by a HiPSTER 1 bipolar power supply from Ionautics. Frequency (4250 - 7500 Hz) and pulse width (10 µs) were chosen based on previous results [43,44], leading to peak currents of 10 ± 1 A at the beginning of the deposition, which typically dropped after several process hours to 7-8 A or in case of longer target life down to 4 A. Oxide targets were sputtered with radio-frequency magnetron sputtering (RFMS) at powers between 60 and 75 W, the Al/Si alloy target at 50 W. All magnetrons were unbalanced and – except from the first approach – open-field configurations were used. Depositions were typically performed at 5.5 ± 0.2 µbar pressure with a gas composition of 14:6 sccm Ar:$N_2$ (all 99.9999% pure), while argon was supplied at the bottom of the chamber and nitrogen directly to the metallic Al and Si targets. Depending on the approach, additional $O_2$ or 10% $O_2$ in $N_2$ gas mixtures were used (99.999% purity). In order to reduce the $O_2$ fraction in the sputter atmosphere, gas flows of 28:12 sccm Ar:$N_2$ were chosen, when $O_2$ was used (keeping 5.5 ± 0.2 µbar pressure). The Corning® EAGLE XG® substrates (2"x2", 1.1 mm thick) were static and either not intentionally heated or heated to approximately 410°C. Prior to deposition, substrates were cleaned for 5 min in acetone and ethanol in an ultrasonic bath. The substrate was sputter cleaned by applying 100 V RF bias (10-11 W) before deposition. The pre-sputtering target cleaning was performed for 26 min at 5.5 µbar in Ar and the target poisoning procedure was held for 4 min at 5.5 µbar with 14:6 sccm Ar:$N_2$. Applying an RF bias to the substrate during deposition lead to significant Ar incorporation into the films (cf. **Fig. S4**), leading to partial delamination above a certain film thickness. Therefore, the presented sample libraries were deposited without the application of a substrate bias (grounded holder, floating glass substrate). To obtain sufficiently thick films for the nanoindentation measurements, the deposition times varied between 6 and 16 h. The base pressure was always below $4 \cdot 10^{-8}$ mbar.

## 2.2. Characterization

High-throughput characterization techniques were applied after the deposition on 45 pre-defined sample points on each materials library: Starting with X-ray photoelectron spectroscopy (XPS) mapping, followed by X-ray diffraction (XRD), UV-Vis-NIR spectrophotometry and nanoindentation. Compositional analysis was performed by X-ray photoelectron spectroscopy on a PHI Quantera system after 10 min 2 kV $Ar^+$ sputter cleaning to remove the surface oxide layer. Monochromatized Al K$\alpha_1$ ration was used for excitation, while the anode was operated at 50 W and 15 kV with a spot size of 200 µm. A dual beam neutralizer (electrons and $Ar^+$ ions) was used to mitigate charging issues. Compositions were determined after Shirley background subtraction (with an additional offset in the case of Si 2p) using the CasaXPS software. To calibrate the anion composition obtained after sputtering, Heavy Ion ToF-ERDA (HI-ERDA) measurements were performed with a 13 MeV iodine beam under a total scattering angle of 36° on selected samples at the Laboratory of Ion Beam Physics at ETH Zurich [45]. The calibration curve and exemplary ERDA results are displayed in **Fig. S2**. The specific procedure for samples analyzed via UHV transfer between deposition chamber and XPS can be found in the Supplementary Information.

Structural analysis was performed by X-ray diffraction (XRD) on a Bruker D8 Discover equipped with a 4-circle goniometer and Cu K$\alpha$ source. Transmission electron microscopy (TEM) was performed using a JEM-2200FS JEOL electron microscope operated at 200 keV. The cross-sectional TEM samples were prepared by using the lift-out method in a dual-beam focused ion beam (a FEI Helios NanoLab G3 UC Dual Beam SEM/$Ga^+$ FIB system). The final TEM sample thinning was done at an ion current of 24 pA at 5 kV to a thickness below 100 nm.

UV-Vis-NIR spectrophotometry measurements were executed on a home-built setup in the range of 200-1100 nm. Detailed information on the setup can be found in [46]. Refractive indices, thickness and absorption coefficients were determined from the transmission spectra using Swanepoel's envelope method [47–49] implemented in a MATLAB® script. More information on the code is provided in the Supplementary Information.

Nanoindentation mapping was performed under load control using a Fischer Picodentor HM 500 equipped with a diamond Berkovich indenter. Indentation hardness and reduced Young's modulus were determined using the Oliver-Pharr method [50], fitting the unloading curve with a polynomial between 20-98% of the load. Nine load-controlled indents were executed on each sample point, 20 µm apart, to a maximum indentation depth of 150 nm (contact depth ≈ 130 nm). The procedure for determining the maximum indentation depth is described in the Supplementary Information.

## 3. Results and discussion

### 3.1 Approaches for O/N-gradients

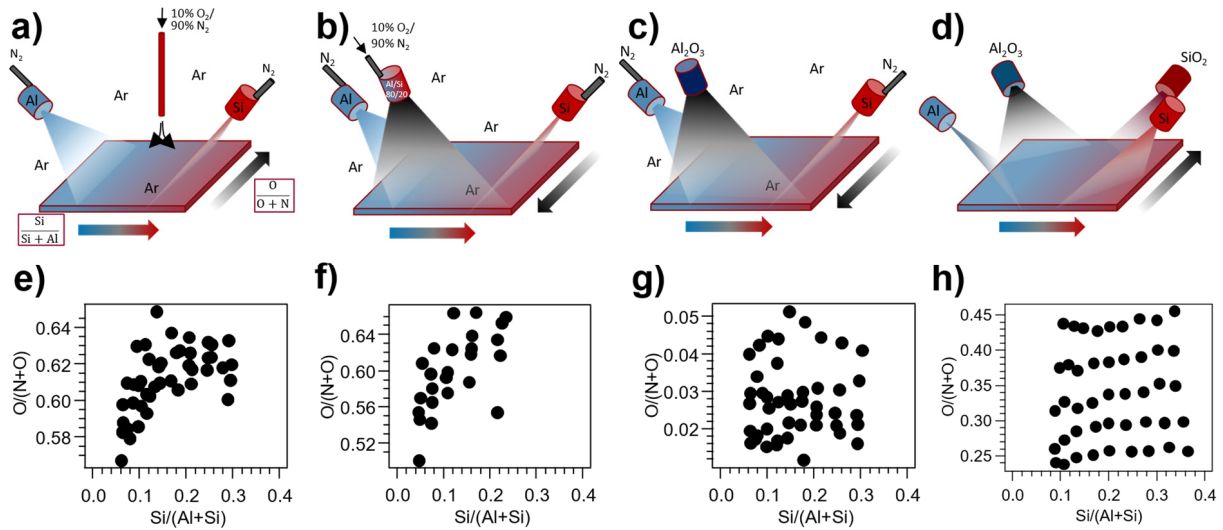

**Fig. 1.** Different approaches to obtain O/N gradients on combinatorial libraries perpendicular to a cation gradient: Deposition setups (a-d) and exemplary anion and cation ratios for a single materials library as measured by XPS for the corresponding setups (e-h).

Error! Reference source not found. shows three different approaches to control anion gradients perpendicular to a cation gradient. The basic setup consists of two metallic targets, aluminum and silicon, which were oriented opposite to each other in a confocal sputter-up geometry, while reactively sputtering in an Ar/$N_2$ atmosphere without substrate rotation. This produced a gradient in the cation concentration between typically ~5% and 30 or 45% Si/(Al+Si) depending on the specific target powers employed. As a first approach, a local $O_2$ gas inlet via a nozzle was installed close to one edge of the substrate to obtain a gradient in the O/N ratio perpendicular to the cation gradient (see **Fig. 1a** for a schematic sketch and **Fig. S5a** for a photograph). $N_2$ on the other hand is directed to the sputter guns. The basic idea behind this setup is that the local gas inlet would produce a variation in oxygen partial pressure across the library, while avoiding oxygen poisoning of the Al and Si targets. Consequently, arriving metal species (atoms or ions) would react at the substrate surface with varying ratios of oxygen and nitrogen species during film growth. The concept of this setup with separated gas inlets was inspired by the work of Fischer *et al.*, who aimed to stabilize the deposition process of Al-O-N [51], and Han *et al.*, who varied the N-doping in MnTe over a combinatorial library via an asymmetric gas inlet [28]. For the second approach (**Fig. 1b**), the O/N gradient was to be obtained by sputtered oxygen species from a third, oxygen-poisoned target. Therefore, $O_2$ containing gas was injected directly to an alloy target with Al/Si = 80/20 composition. This target composition was selected to reduce the influence on the cation gradient. For the third approach (**Fig. 1c**), an oxide target was used instead of $O_2$-containing gas. This can be compared to the use of differently oxidized precursors in MBE [30] or the use of oxide and nitride compound targets in PLD [33]. To reduce the effect of additional temperature gradients due to plasma heating or a gradient in Ar-bombardment across the library, the magnetic configuration requires attention. For the setups with more than two magnetrons, the open-field configuration was chosen as the most symmetrical option.

Below the schematic deposition schemes, representative anion and cation spreads measured by XPS are displayed in **Fig. 1e-g**. For both approaches using $O_2$ gas, the anion and cation gradients are not independent, as the approaches lead to anion gradients with a significant contribution in the same direction as the cation gradient. Furthermore, the obtained relative spread in the O/N ratios per library is comparably small (less than 15% per library). In both cases partial oxygen poisoning of the targets contributes to an increase in the overall oxygen content, while a different poisoning behavior of the silicon and the aluminum targets further convolutes the cation and anion gradients. This is consistent with an earlier onset of oxygen poisoning of the silicon target compared to aluminum as observed in hysteresis experiments (c.f. **Fig. S6a-c**). A second contributing factor could be the different affinities of Al and Si species to react with oxygen at the substrate surface. The formation enthalpies of both oxides and nitrides are collected in **Tab. S1** [52]. Silicon exhibits oxide formation enthalpies which are ~3.7 times higher than those of the nitride, while this factor is only ~2.6 in the case of aluminum, which could lead to the higher O/(N+O) ratios in the silicon-rich samples.

The third approach – avoiding $O_2$ gas supply – exhibits independent anion and cation gradients (see **Fig. 1g**). Sputtering from the oxide target presumably supplies a directional flux of oxygen species apparently without the limitation of a high background partial pressure of oxygen in the vacuum chamber. Although this approach results in reproducible O/N gradients, the three-magnetron setup presented has limitations when higher oxygen levels are of interest, since the only oxygen source is also an aluminum source. To mitigate this limitation, the third approach can be extended by adding a second oxide target as depicted in **Fig. 1d**, leading to orthogonal anion and cation gradients (compare **Fig. 1h**). In addition, comparably large spreads of the O/N ratio can be achieved (more than 40% per 2"x2" library). Allowing for the best control to obtain independent anion and cation gradients, the third approach and its extension were applied to study the Al-Si-O-N thin film system in detail, which is presented in the following sections.

## 3.2 2D-screening of the Al-Si-O-N system

### 3.2.1 Structural screening

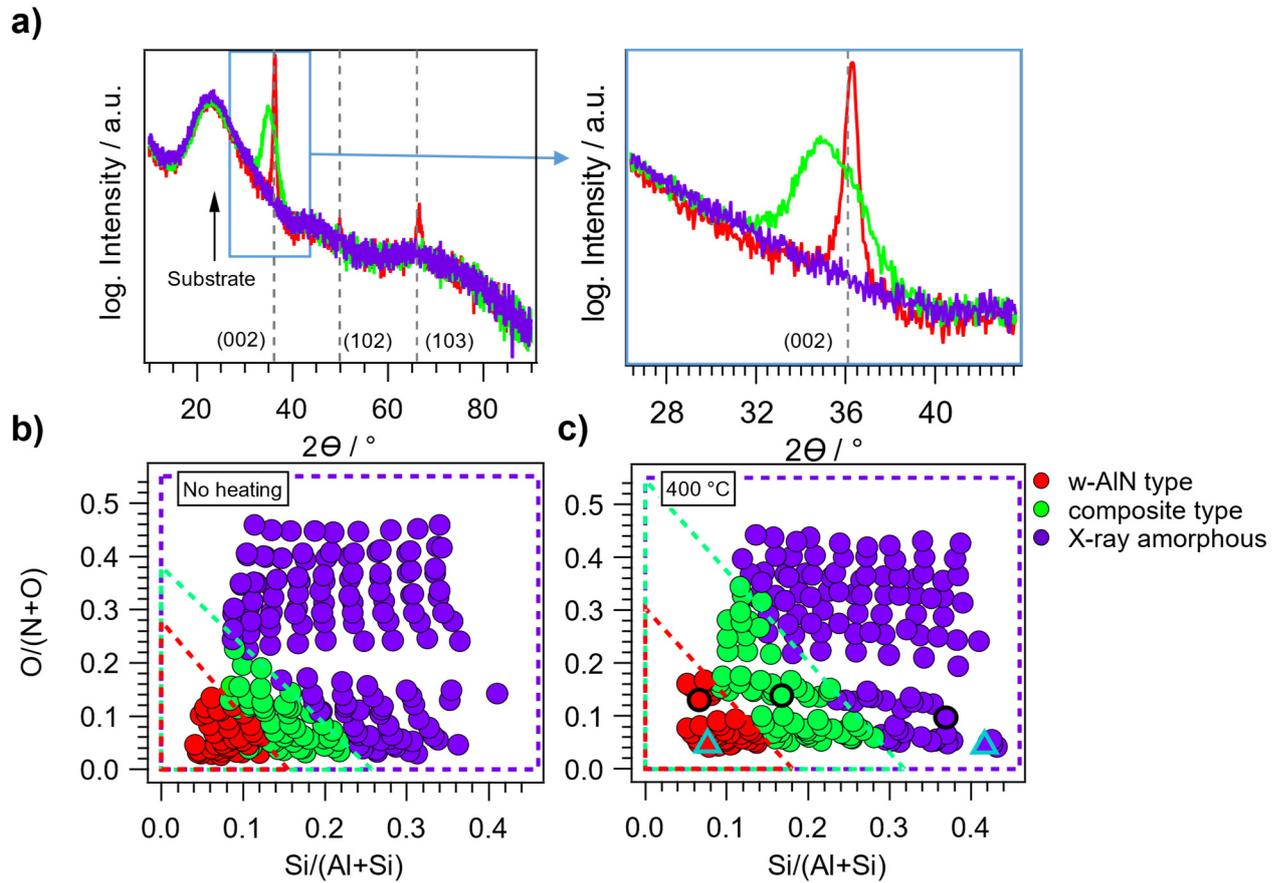

**Fig. 2.** Structural analysis based on automated XRD mapping: a) Representative diffraction patterns for the three observed structural types based on wurtzite-AlN (red), a composite structure (green) and X-ray amorphous samples (violet); cation-anion phase diagram for samples deposited b) without external heating and c) at 400 °C. The highlighted samples are those presented in a) (O) and those investigated by TEM (Δ). A pattern was classified as composite type, if a second pseudo-Voigt peak is required to obtain a decent fit of the (002) reflection, and as amorphous type, if the pattern coincides with the one of the bare glass substrate. Indicated reflections are based on powder diffraction of AlN reported by Schulz and Thiemann [53].

Applying the third approach and its extension introduced in the previous section, a cation range of 4–44% Si/(Al+Si) and an anion range of 2-46% O/(N+O) was screened by five depositions. The depositions were performed both at 400 °C and without external heating (see **Fig. S7**). A first structural characterization was carried out using automated XRD mapping (see **Fig. 2**). **Fig. 2b,c** summarize the results of the phase screening. Three different characteristic types of XRD patterns were observed: The first can be assigned to wurtzite AlN (w-AlN) featuring (002), (103) and (102) reflections (red pattern in **Fig. 2a**). This phase has been identified at low silicon and oxygen contents (red markers in **Fig. 2b,c**). The second type shows a shoulder or a broad peak at lower

2Θ angles adjacent to the w-AlN (002) reflection (green pattern in **Fig. 2a**). Simultaneously, the w-AlN reflection intensities decrease with increasing Si/(Al+Si) and O/(N+O) ratios. The broad hump has been attributed in the literature to lattice plane dilatation at grain boundaries or a partially ordered grain boundary phase [38]. A comparison with diffraction patterns from the ICSD database reveals that α-$Si_3N_4$ [54], α-$Al_2O_3$ [55], $SiAl_4O_2N_4$ [56] or mullite $Al_{6(1-z)}Si_{6z}O_{9+3z}$ [57] could be possible candidates for the latter, exhibiting reflections in the respective region. Therefore, this phase type will be referred to as the "composite" phase region in the following and is found at intermediate Si and O contents. The third type represents peak-free patterns indicating an X-ray amorphous microstructure (purple pattern in **Fig. 2a**) and occurs at even higher oxygen and silicon concentrations. Two main observations can be made based on the cation-anion phase diagrams in **Fig. 2b,c**: Firstly, there is an apparent linear relationship between the Si/(Al+Si) and O/(N+O) fractions on the boundaries separating the three phases. Secondly, a slight expansion of the "w-AlN"-type region and a significant expansion of the "composite" region along the Si/(Si+Al) and O/(N+O) axes could be achieved by increasing the deposition temperature.

A similar microstructural evolution is reported in the literature for the ternary systems Al-Si-N and Al-O-N. Four regions are differentiated [11,37,38]: A solid solution of Si or O in w-AlN is followed by a nanocomposite of w-AlN grains surrounded by a thin grain boundary phase. A further increase in Si or O content leads to a thickening of the grain boundary phase, finally ending in an amorphous phase. In the present study, the observed shift of the (002) peak position in the XRD patterns indicate a Si solubility limit within the "w-AlN"-type region (c.f. **Fig. S8**), Therefore, the "w-AlN"-type region (red) likely encompasses both a solid solution and a first nanocomposite regime with a grain boundary phase not detectable by the XRD mapping procedure. Within the subsequent "composite" phase (green), the presumably partially crystalline grain boundary phase thickens, while the volume fraction of w-AlN based grains decreases until the grain size and their amount are reduced to a level not detectable by XRD (purple). Increasing the substrate temperature may promote the crystallization of w-AlN grains and in particular the nucleation of the grain boundary phase resulting in the observed significant extension of the composite region at 400 °C.

To assess the microstructural features of the different phase regions in more detail, TEM microscopy micrographs were obtained for two samples with low O/(N+O) ratio (3.5%) and 400 °C deposition temperature. **Fig. 3a** shows a bright-field TEM (BF-TEM) cross-sectional image of the sample with 8% Si/(Al+Si) that belongs to the "w-AlN"-type region. The film exhibits a dense microstructure composed of w-AlN columnar grains. The columnar grains are tilted by about +10° with respect to the substrate normal because the substrate was not rotated using a confocal sputter geometry. This phenomenon is known as the tangent rule and occurs due to shadowing and slow (surface) diffusion at low substrate temperatures [58,59]. The corresponding selected area electron diffraction (SAED) pattern (**Fig. 3a** insert) reveals a series of diffraction rings and arcs associated with w-AlN implying a polycrystalline microstructure. Importantly, the {002} reflections have the highest intensity with relatively narrow angular distribution representing an arc-like diffraction intensity. The {002} reflections of the highest intensity are inclined from the film growth direction implying an abnormal (002) texture. A high-resolution TEM (HRTEM) analysis (**Fig. 3b**) reveals coherent lattice fringes along the [002] direction within the grain interior which are inclined by approximately 80° from the grain growth direction. This finding stands in contrast to literature reports on the ternary subsystems [11,37], as well as on pure AlN [43], in which the majority of the w-AlN (002) planes are oriented along the grain growth direction. The potential origin of this phenomenon is currently the subject of further investigation. In **Fig. 3b,** a BF-TEM cross-sectional image of the sample with 42% Si/(Al+Si), which belongs to the "X-ray amorphous" region, is displayed showing a featureless microstructure. The corresponding SAED patterns exhibit broad diffraction rings typical for amorphous materials. However, the HRTEM image reveals the presence of small (2-5 nm diameter), randomly oriented equiaxed w-AlN grains surrounded by an amorphous matrix. Due to their small size, these grains were not detected by the XRD mapping. Similar findings have been reported by Pélisson *et al.* [37], Musil *et al.* [40] and Fischer *et al.* [11].

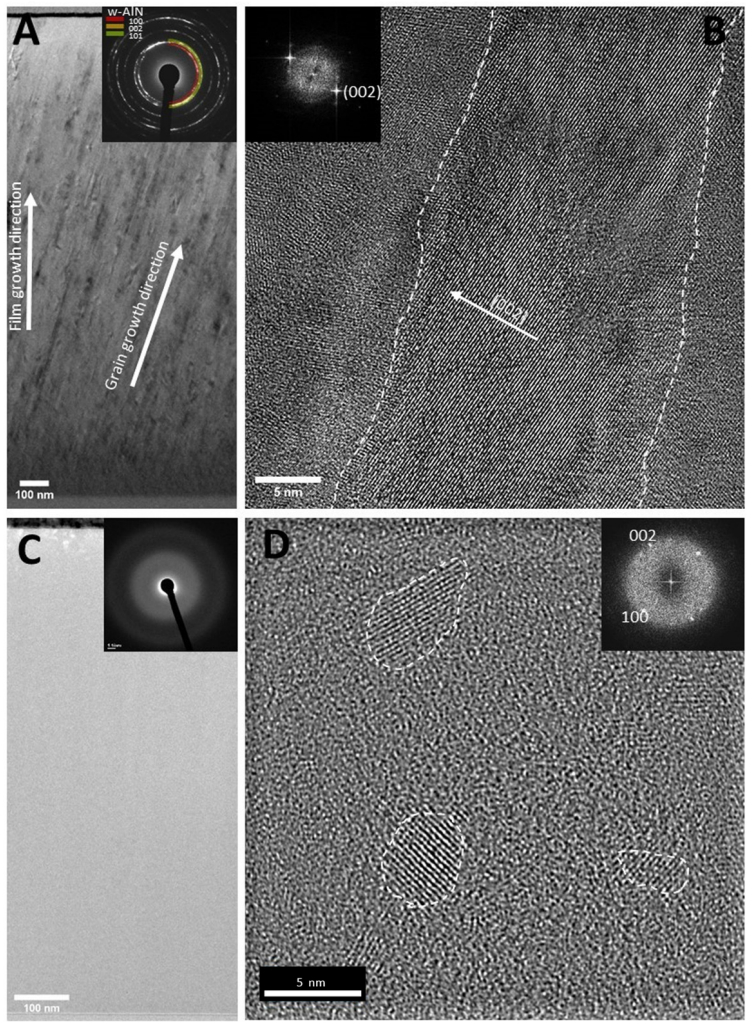

**Fig. 3.** Low-magnification BF-TEM images and corresponding SAED patterns as insets from the sample (a) with 8% Si/(Al+Si) and (c) 42% Si/(Al+Si) and 3.5% O/(N+O) grown at 400 °C. High-resolution BF-TEM cross-section images with corresponding fast-Fourier Transform (FFT) images as insets for the sample (b) with 8% Si/(Al+Si) and (d) 42% Si/(Al+Si) and 3.5% O/(N+O) grown at 400 °C (see **Δ** in **Fig. 2d**).

### 3.2.2. Functional property screening

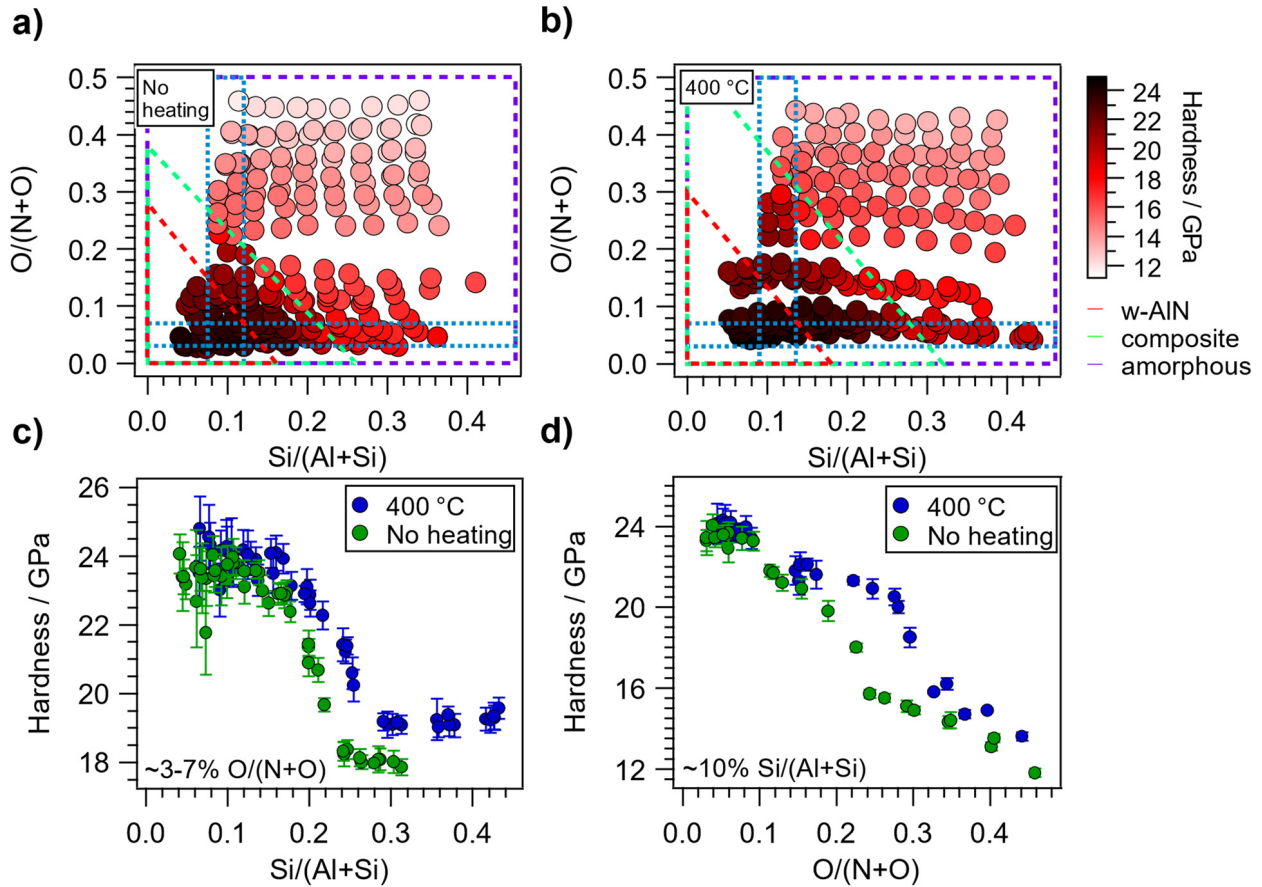

**Fig. 4.** Nanoindentation mapping results for Al-Si-O-N thin film libraries: Hardness as a function of the anion and cation composition for libraries deposited a) without external heating and b) at 400 °C. The dashed lines indicate the presence of the structural types observed in **Fig. 2**. Cross sections of hardness versus c) cation composition at low oxygen content and versus d) anion composition at low silicon content. Markers denote the mean; error bars indicate the standard deviation over nine indents.

The mechanical properties of the Al-Si-O-N thin film libraries were screened using automated nanoindentation mapping. In **Fig. 4**, the resulting hardness values are displayed as a function of the anion and cation composition. A clear correlation with the microstructural evolution discussed in the previous section is evident. A high hardness plateau of 23-25 GPa is present in the region with "w-AlN"-type. Within the "composite" region, the hardness decays with increasing silicon and oxygen content, starting slowly followed by a steep drop close to the transition to the "X-ray amorphous" region. The prominent decrease in hardness at the end of the "composite" phase has been observed in several nanocomposite systems and has been attributed to the thickening of the soft grain boundary phase [60,61]. Due to the extension of the crystalline region at elevated temperatures, this hardness decrease is shifted towards higher Si and O contents for 400 °C deposition temperature. In the "X-ray amorphous" region, a second plateau is present with hardness values of 18.1 ± 0.2 GPa (no heating) or 19.3 ± 0.3 GPa (400 °C) for samples with low oxygen contents of 3-7% O/(N+O) (see **Fig. 4c**). This observation could be related to the presence of nanocrystals in the amorphous matrix as identified in the TEM micrographs. Their growth could be favored by faster diffusion processes at higher deposition temperatures. In addition to the observed hardness drop associated with the different microstructure types, an almost linear decrease in hardness is observed with increasing oxygen contents as visualized in **Fig. 4d**. We hypothesize that a higher degree of iconicity of Al-O bonds compared to Al-N bonds contributes to this effect, resulting in lower elastic moduli and lower hardness [62].

The measured hardness values in the "w-AlN"-type and "X-ray amorphous" regions are in good agreement with those reported by Fischer *et al.* [11] and Lewin *et al.* [42]. Contrary to findings for other nanocomposite systems, the present material screening does not reveal a substantial increase in hardness for intermediate Si contents, where the hardening due to the "strongest size" effect could have been expected due to a nanocomposite microstructure [62–64]. On the one hand, the applied process conditions and oxygen contents > 1% O/(N+O) may not result in the required microstructure of 10-15 nm sized w-AlN grains embedded in a monolayer thick $SiN_x$ grain boundary phase with a sharp interface [65–67]. On the other hand, theoretical calculations by Sheng *et al.* suggest that spinodal decomposition is improbable in the Al-Si-N system, which hinders the formation of the microstructure required for a strong hardness enhancement, and could therefore represent a general limitation in this material system [68].

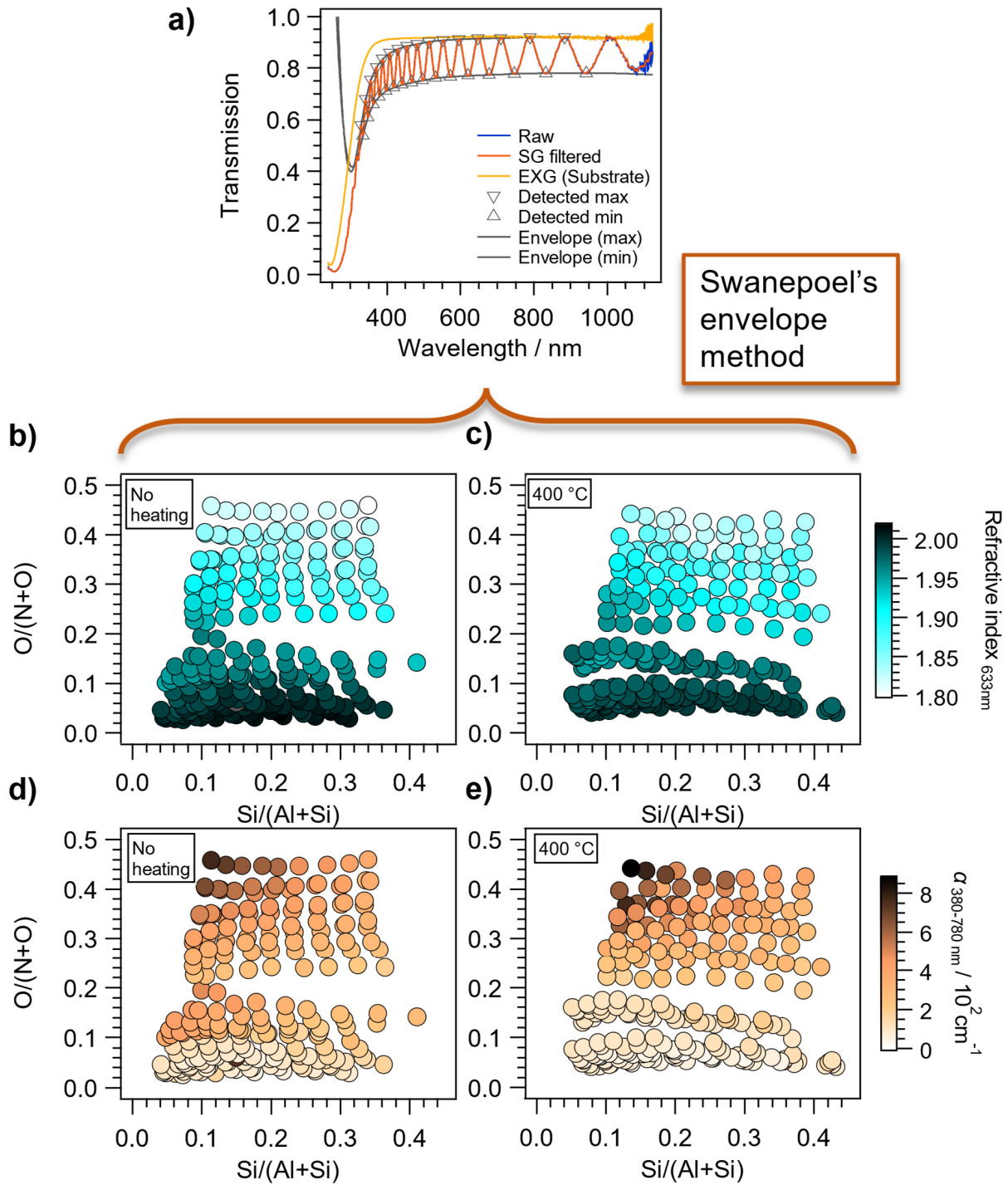

**Fig. 5.** Optical property screening for Al-Si-O-N thin film libraries based on UV-Vis transmission data and Swanepoel's envelope method: a) Typical transmission spectrum of the sample on the EXG substrate, the substrate as well as determined features necessary for the Swanepoel analysis. Refractive index at 633 nm (b & c) and the mean absorption coefficient between 380 and 780 nm (d & e) as a function of anion and cation composition derived via the Swanepoel method.

To assess the optical constants in a quick and automated manner, the Swanepoel method was implemented to extract the desired quantities from UV-Vis-NIR transmission spectra. The required features determined from the interference extrema of the transmission spectra are visualized in **Fig. 5a**, while the detailed procedure is described in the Supplementary Information. In contrast to the hardness, a correlation between the refractive index at 633 nm and the determined structure types is barely perceptible (see **Fig. 5Error! Reference source not found.b** and **Fig. 5c**). Therefore, the deposition temperature has no significant influence on the results as well. Instead, the refractive index decreases approximately linearly with increasing O/(N+O) ratio from 2.02 to 1.80 as expected from literature reports [11,41,69,70]. According to Shannon and Fischer, the dynamic polarizability of the fourfold coordinated $Al^{3+}$ cation (0.533 $Å^3$) exceeds that of $Si^{4+}$ (0.284 $Å^3$) [71]. Therefore, a decrease in the refractive index is expected with increasing silicon cation substitution, which can be inferred from the present data for higher oxygen contents. However, the extent of this variation is significantly lower than that due to the O/(N+O) ratio. Indeed, the anion polarizability is significantly higher and has a stronger influence on the refractive index. As presented in the Supplementary Information, the application of the Anderson-Eggleton equation as given by Shannon and Fischer [71] enables the calculation of the nitrogen polarizability in the visible range based on the refractive index and the sample composition. Values between 2.4 and 2.8 $Å^3$ are estimated, while decreasing polarizabilities are found for increasing oxygen contents, particularly pronounced in the "X-ray amorphous" region (c.f. **Fig. S9**). This observation may be attributed to a reduced charge density around the nitrogen ions with more electronegative oxygen ions in close proximity. In order to obtain values with reduced scatter that could serve for predictive purposes, the implementation of a more sophisticated model – considering, for example, a molar volume dependence or a varying degree of covalency – may be required.

All deposited thin films were transparent. Considering the mean absorption coefficient in the visible region, the deposition temperature exhibits only a small influence, while a significant increase is observed at higher oxygen contents (as reported in literature [70]), especially for Al-rich samples. The highest contribution to the absorption originates from wavelength range < 500 nm, as visualized by the maps of the absorption coefficient at different wavelengths in **Fig. S10**. Placing the libraries in front of a white background therefore allows the observation of a slight yellowish appearance for Al- and O-rich regions (see **Fig. S11**). The increased absorption could arise due to scattering at inclusions or precipitates [72]. Their coherent size may be below the inherent detection limit of the XRD mapping.

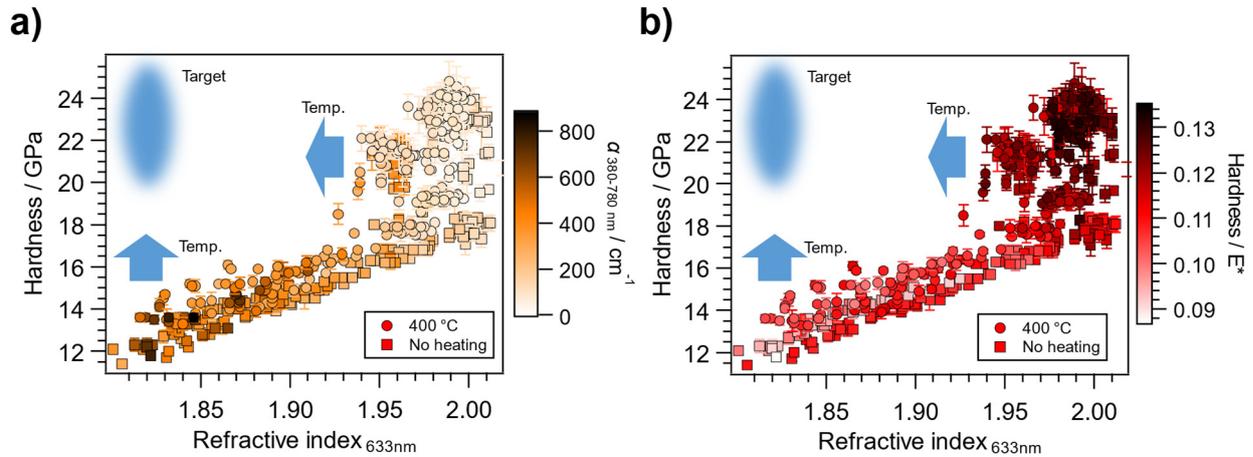

**Fig. 6.** Hardness (*H*) plotted against refractive index (*n*) with different marker color codes for Al-Si-O-N thin film libraries: a) Mean absorption coefficient between 380 and 780 nm, and b) Hardness / effective Young's modulus (*H/E\**) ratio. Blue spots indicate the desired target area.

As mentioned above, hard, resilient coatings with tunable refractive indices are of particular importance in the field of protective anti-reflective coatings. To visualize the results of the co-optimization, hardness H is plotted against the refractive index in **Fig. 6**. There is a general linear trend between hardness and refractive index, particularly for low refractive indices where exclusively the "X-ray amorphous" regime prevails. The independent tunability of hardness and refractive index is limited to a refractive index range between 1.94 and 2.02 within the screened compositional space of the "w-AlN"-type and the "composite" microstructure. When the deposition temperature is increased to 400 °C, the hardness in the "X-ray amorphous" region is raised by approximately 1 GPa for the same refractive indices. In addition, the variable hardness range is extended towards slightly lower refractive indices. Considering the extent of the observed hardness increase, the likelihood of achieving combinations of properties in the target area (blue spots in **Fig. 6**) appears low, even with a possible further increase in deposition temperature.

Another potential route to extend the "composite" regime towards higher oxygen contents is the reduction of the silicon content, as illustrated by Error! Reference source not found.. This could enable increased hardness also for lower refractive indices than 1.94. Nonetheless, a linear decrease with the refractive index is evident for the high hardness limit as well, representing an additional impediment to accessing the target area of > 20 GPa.

To assess the origin of the apparent linear correlation between hardness and refractive index, possible common descriptors are required. One possible descriptor could be a measure for the degree of bond iconicity or covalency. As described in the Supplementary Information, the XPS binding energy difference between anion and cations can represent a measure for bond covalency. Following these considerations, **Fig. S13** illustrates that refractive index, hardness and the effective Young's modulus correlate only partially to the material's covalency. A more comprehensive discussion can be found in the Supplementary Information.

In addition to high hardness and a suitable refractive index, a high level of transparency is desirable for the application as an anti-reflective coating. **Fig. 6a** depicts the mean absorption coefficient in the visible spectrum as a color code showing low values of ≤ $10^2$ cm$^{-1}$ for samples with the highest refractive indices and hardness. In the context of anti-reflection coatings on flexible substrates, the coating must demonstrate resistance to cracking during bending additionally. Musil *et al.* have found that the ratio between hardness and effective Young's modulus ($H/E^*$) can serve as an estimate for the resistance to cracking due to substrate bending. In their studies, no cracking was observed for coating materials with $H/E^* \geq 0.1$ [73]. As visualized in **Fig. 6b**, most of the Al-Si-O-N samples fulfil this criterion, even for lower refractive indices. The highest values of $H/E^* \geq 0.13$ are again present in high hardness and high refractive index samples with "w-AlN"-type microstructure.

## 4. Conclusions

In this study we performed a high-throughput combinatorial investigation of Al-Si-O-N thin films, exploring both cation and anion composition variations. Among all tested approaches, we determined that the combination of metallic (Al, Si) and oxide (SiO$_2$ and Al$_2$O$_3$) targets allowed us to achieve distinct orthogonal cation and anion gradients and a decent spread in the oxygen/nitrogen ratio across the material library. This approach enabled the screening of a compositional space of approximately 40% in the Si/(Al+Si) ratio and 45% in the O/(N+O) ratio synthesizing only five thin film libraries. This time-efficient screening allows for an additional investigation of the influence of process parameters – in this study the deposition temperature – on material properties of interest. We determined three composition-dependent types of microstructures: a "wurtzite-AlN"-based microstructure for low silicon and oxygen contents, followed by a "composite" microstructure and an X-ray amorphous microstructure for higher silicon and oxygen contents. The measured hardness followed these structural zones, starting with a high hardness plateau of 23-25 GPa for the "wurtzite AlN"-type phase, undergoing a drop at the transition between the "composite" and the "X-ray amorphous" phase, and reaching another hardness plateau of 18-19 GPa at low oxygen contents. An increase in the oxygen content superposes a further linear decrease in hardness down to 11.5 GPa. The plateau of high hardness could be extended by increasing the deposition temperature to 400 °C with 5% Si/(Al+Si) and 8% O/(N+O). In contrast, in agreement with the previous reports, the refractive index is mainly influenced by the oxygen/nitrogen ratio and to a lesser extent by the cation ratio, resulting in a span between 1.80 and 2.02 within the screened composition space. Our results illustrate the linear correlation between refractive index and hardness, particularly within the "X-ray amorphous" region. The higher hardness phase, which allows this linear relationship to be overcome, is strongly constrained in the compositional space. Consequently, the ability to tune the refractive index yet keeping high hardness values is also significantly limited in this material system.

With this proof-of-principle, we hope to encourage other researchers to adopt the herein presented methodology for orthogonal anion and cation gradients to other compositionally complex systems to accelerate development in many other application fields.

**Acknowledgments**

S. F. acknowledges funding by the Helmut Fischer und Anni Walther Stiftung. O.P. acknowledges funding from the SNF (projects 227945). A.W. and S.S. acknowledge funding from the Strategic Focus Area – Advanced Manufacturing (SFA-AM) through the project Advanced manufacturability of hybrid organic-inorganic semiconductors for large area optoelectronics (AMYS). Sasa Vranjkovic is gratefully acknowledged for technical support with the custom adaptations of the deposition chamber.

**Authors' contributions**

**S. F.:** Conceptualization, Investigation, Formal analysis, Visualization, Writing - Original Draft; **O.P.:** Investigation, Formal Analysis; Writing – Review & Editing; **A.M.:** Investigation, Formal Analysis, Writing – Review & Editing; **A.W.:** Investigation, Writing – Review & Editing; **K.T.:** Investigation, Writing – Review & Editing; **S.S.:** Conceptualization, Supervision, Methodology, Formal analysis, Funding acquisition, Writing – Review & Editing

## 5. Bibliography

[1] M. Koyama et al., *Effects of Nitrogen in HfSiON Gate Dielectric on the Electrical and Thermal Characteristics*, in *Digest. International Electron Devices Meeting,* (IEEE, 2002), pp. 849–852.

[2] L.-P. Feng and Z.-T. Liu, *HfSiON Films Deposited by Radio Frequency Reactive Sputtering*, in *Metallic Oxynitride Thin Films by Reactive Sputtering and Related Deposition Methods: Process, Properties and Applications* (2013), pp. 230–253.

[3] Y.-H. Wu and J.-H. Huang, *Application of Oxynitrides for Microelectronic Device and Gas Barriers*, in *Metallic Oxynitride Thin Films by Reactive Sputtering and Related Deposition Methods: Process, Properties and Applications*, edited by F. Vaz, N. Martin, and M. Fenker (Bentham Science Publisher, 2013), pp. 285–339.

[4] G. Hitoki, T. Takata, J. N. Kondo, M. Hara, H. Kobayashi, and K. Domen, An oxynitride, TaON, as an efficient water oxidation photocatalyst under visible light irradiation (λ ≤ 500 nm), Chem. Commun. **2**, 1698 (2002).

[5] R. Xie and H. T. (Bert) Hintzen, Optical Properties of (Oxy)Nitride Materials: A Review, J. Am. Ceram. Soc. **96**, 665 (2013).

[6] Y.-I. Kim, Syntheses, Crystal Structures, and Dielectric Property of Oxynitride Perovskites, Ohio State University, 2005.

[7] S. Yamada, H. Emoto, M. Ibukiyama, and N. Hirosaki, Properties of SiAlON powder phosphors for white LEDs, J. Eur. Ceram. Soc. **32**, 1355 (2012).

[8] L. Rebouta, A. Sousa, M. Andritschky, F. Cerqueira, C. J. Tavares, P. Santilli, and K. Pischow, Solar selective absorbing coatings based on AlSiN/AlSiON/AlSiO$_y$ layers, Appl. Surf. Sci. **356**, 203 (2015).

[9] Y. Wang, X. Cheng, Z. Lin, C. Zhang, and F. Zhang, Optimization of PECVD silicon oxynitride films for anti-reflection coating, Vacuum **72**, 345 (2003).

[10] M. Urgen, V. Ezirmik, E. Senel, Z. Kahraman, and K. Kazmanli, The effect of oxygen content on the temperature dependent tribological behavior of Cr–O–N coatings, Surf. Coatings Technol. **203**, 2272 (2009).

[11] M. Fischer, M. Trant, K. Thorwarth, R. Crockett, J. Patscheider, and H. J. Hug, Understanding the microstructural evolution and mechanical properties of transparent Al-O-N and Al-Si-O-N films, Sci.

Technol. Adv. Mater. **20**, 1031 (2019).

[12] M.-C. Lin, L.-S. Chang, and H. C. Lin, Gas barrier properties of titanium oxynitride films deposited on polyethylene terephthalate substrates by reactive magnetron sputtering, Appl. Surf. Sci. **254**, 3509 (2008).

[13] P. Carvalho, L. Cunha, N. P. Barradas, E. Alves, J. P. Espinos, and F. Vaz, *Tuneable Properties of Zirconium Oxynitride Thin Films*, in *Metallic Oxynitrid Thin Films by Reactive Sputtering and Related Deposition Methods*, edited by F. Vaz, N. Martin, and M. Fenker (Bentham Science Publishers Ltd., 2013), pp. 64–112.

[14] J. Vetter, *Oxynitrides and Oxides Deposited by Cathodic Vacuum Arc*, in *Metallic Oxynitride Thin Films by Reactive Sputtering and Related Deposition Methods: Process, Properties and Applications* (2013), pp. 265–284.

[15] G. J. Wan, N. Huang, S. C. H. Kwok, Z. Y. Shao, A. S. Zhao, P. Yang, and P. K. Chu, Si–N–O Films Synthesized by Plasma Immersion Ion Implantation and Deposition (PIII&D) for Blood-Contacting Biomedical Applications, IEEE Trans. Plasma Sci. **34**, 1160 (2006).

[16] M. Fenker, *Properties of Oxynitride Thin Films for Biomedical Applications*, in *Metallic Oxynitride Thin Films by Reactive Sputtering and Related Deposition Methods: Process, Properties and Applications* (2013), pp. 254–264.

[17] D. Heřman, J. Šícha, and J. Musil, Magnetron sputtering of TiO N films, Vacuum **81**, 285 (2006).

[18] H. Akazawa, Transparent conductive properties of TiON thin films, J. Vac. Sci. Technol. A **40**, (2022).

[19] K. Alberi et al., The 2019 materials by design roadmap, J. Phys. D. Appl. Phys. **52**, 013001 (2019).

[20] M. Stueber, D. Diechle, H. Leiste, and S. Ulrich, Synthesis of Al–Cr–O–N thin films in corundum and f.c.c. structure by reactive r.f. magnetron sputtering, Thin Solid Films **519**, 4025 (2011).

[21] D. Diechle, M. Stueber, H. Leiste, and S. Ulrich, Combinatorial approach to the growth of α-$(Al_{1-x},Cr_x)_{2+\delta}(O_{1-y},N_y)_3$ solid solution strengthened thin films by reactive r.f. magnetron sputtering, Surf. Coatings Technol. **206**, 1545 (2011).

[22] S. Spitz, M. Stueber, H. Leiste, S. Ulrich, and H. J. Seifert, Phase formation and microstructure evolution of reactively r.f. magnetron sputtered Cr–Zr oxynitride thin films, Surf. Coatings Technol. **237**, 149 (2013).

[23] C. L. Melamed, M. B. Tellekamp, J. S. Mangum, J. D. Perkins, P. Dippo, E. S. Toberer, and A. C. Tamboli, Blue-green emission from epitaxial yet cation-disordered $ZnGeN_{2-x}O_x$, Phys. Rev. Mater. **3**, 051602 (2019).

[24] S. K. Suram, S. W. Fackler, L. Zhou, A. T. N'Diaye, W. S. Drisdell, J. Yano, and J. M. Gregoire, Combinatorial Discovery of Lanthanum–Tantalum Oxynitride Solar Light Absorbers with Dilute Nitrogen for Solar Fuel Applications, ACS Comb. Sci. **20**, 26 (2018).

[25] K. R. Talley, J. Mangum, C. L. Perkins, R. Woods-Robinson, A. Mehta, B. P. Gorman, G. L. Brennecka, and A. Zakutayev, Synthesis of Lanthanum Tungsten Oxynitride Perovskite Thin Films, Adv. Electron. Mater. **5**, (2019).

[26] S. Siol, A. Holder, B. R. Ortiz, P. A. Parilla, E. Toberer, S. Lany, and A. Zakutayev, Solubility limits in quaternary SnTe-based alloys, RSC Adv. **7**, 24747 (2017).

[27] S. Siol et al., Negative-pressure polymorphs made by heterostructural alloying, Sci. Adv. **4**, eaaq1442 (2018).

[28] Y. Han, B. Matthews, D. Roberts, K. R. Talley, S. R. Bauers, C. Perkins, Q. Zhang, and A. Zakutayev, Combinatorial Nitrogen Gradients in Sputtered Thin Films, ACS Comb. Sci. **20**, 436 (2018).

[29] J. Park, S. H. Shin, J.-S. Bae, X. Zhang, I. Takeuchi, and S. Lee, Combinatorial synthesis of non-stoichiometric $SiO_x$ thin films via high-throughput reactive sputtering, J. Appl. Phys. **129**, (2021).

[30] H.-T. Zhang, L. Zhang, D. Mukherjee, Y.-X. Zheng, R. C. Haislmaier, N. Alem, and R. Engel-Herbert,

Wafer-scale growth of VO$_2$ thin films using a combinatorial approach, Nat. Commun. **6**, 8475 (2015).

[31] G. Hyett and I. P. Parkin, A combinatorial approach to phase synthesis and characterisation in atmospheric pressure chemical vapour deposition, Surf. Coatings Technol. **201**, 8966 (2007).

[32] I. Parkin, A. Kafizas, C. W. Dunnill, and G. Hyett, Combinatorial CVD: New Oxy-nitride Photocatalysts, ECS Trans. **25**, 1239 (2009).

[33] K. V. L. V. Narayanachari, D. B. Buchholz, E. A. Goldfine, J. K. Wenderott, S. M. Haile, and M. J. Bedzyk, Combinatorial Approach for Single-Crystalline TaON Growth: Epitaxial β-TaON (100)/α-Al$_2$O$_3$ (012), ACS Appl. Electron. Mater. **2**, 3571 (2020).

[34] M. Stüber, U. Albers, H. Leiste, K. Seemann, C. Ziebert, and S. Ulrich, Magnetron sputtering of hard Cr–Al–N–O thin films, Surf. Coatings Technol. **203**, 661 (2008).

[35] T. Miyata, Y. Mochizuki, and T. Minami, *Combinatorial Deposition by r.f. Magnetron Sputtering Using Subdivided Powder Targets as New Development Method for Thin-Film Phosphors*, in *Device and Process Technologies for Microelectronics, MEMS, and Photonics IV*, edited by J.-C. Chiao, A. S. Dzurak, C. Jagadish, and D. V. Thiel, Vol. 6037 (2005), p. 60371U.

[36] A. Pélisson, Al-Si-N Transparent Hard Nanostructured Coatings, University Basel, 2009.

[37] A. Pélisson, M. Parlinska-Wojtan, H. J. Hug, and J. Patscheider, Microstructure and mechanical properties of Al–Si–N transparent hard coatings deposited by magnetron sputtering, Surf. Coatings Technol. **202**, 884 (2007).

[38] A. Pélisson-Schecker, H. J. Hug, and J. Patscheider, Morphology, microstructure evolution and optical properties of Al-Si-N nanocomposite coatings, Surf. Coatings Technol. **257**, (2014).

[39] M. Parlinska-Wojtan, A. Pélisson-Schecker, H. J. Hug, B. Rutkowski, and J. Patscheider, AlN/Si$_3$N$_4$ multilayers as an interface model system for Al$_{1-x}$Si$_x$N/Si$_3$N$_4$ nanocomposite thin films, Surf. Coatings Technol. **261**, 418 (2015).

[40] J. Musil, M. Šašek, P. Zeman, R. Čerstvý, D. Heřman, J. G. Han, and V. Šatava, Properties of magnetron sputtered Al–Si–N thin films with a low and high Si content, Surf. Coatings Technol. **202**, 3485 (2008).

[41] A. Bendavid, P. . Martin, and H. Takikawa, The properties of nanocomposite aluminium–silicon based thin films deposited by filtered arc deposition, Thin Solid Films **420–421**, 83 (2002).

[42] E. Lewin, D. Loch, A. Montagne, A. P. Ehiasarian, and J. Patscheider, Comparison of Al–Si–N nanocomposite coatings deposited by HIPIMS and DC magnetron sputtering, Surf. Coatings Technol. **232**, 680 (2013).

[43] J. Patidar, A. Sharma, S. Zhuk, G. Lorenzin, C. Cancellieri, M. F. Sarott, M. Trassin, K. Thorwarth, J. Michler, and S. Siol, Improving the crystallinity and texture of oblique-angle-deposited AlN thin films using reactive synchronized HiPIMS, Surf. Coatings Technol. **468**, 129719 (2023).

[44] J. Patidar, K. Thorwarth, T. Schmitz-Kempen, R. Kessels, and S. Siol, Deposition of highly crystalline AlScN thin films using synchronized high-power impulse magnetron sputtering: From combinatorial screening to piezoelectric devices, Phys. Rev. Mater. **8**, 095001 (2024).

[45] M. Döbeli, C. Kottler, F. Glaus, and M. Suter, ERDA at the low energy limit, Nucl. Instruments Methods Phys. Res. Sect. B Beam Interact. with Mater. Atoms **241**, 428 (2005).

[46] A. Wieczorek, A. G. Kuba, J. Sommerhäuser, L. N. Caceres, C. M. Wolff, and S. Siol, Advancing high-throughput combinatorial aging studies of hybrid perovskite thin films via precise automated characterization methods and machine learning assisted analysis, J. Mater. Chem. A **12**, 7025 (2024).

[47] R. Swanepoel, Determination of the thickness and optical constants of amorphous silicon, J. Phys. E. **16**, 1214 (1983).

[48] J. C. Manifacier, J. Gasiot, and J. P. Fillard, A simple method for the determination of the optical constants n, k and the thickness of a weakly absorbing thin film, J. Phys. E. **9**, 1002 (1976).

[49] D. Poelman and P. F. Smet, Methods for the determination of the optical constants of thin films from

single transmission measurements: a critical review, J. Phys. D. Appl. Phys. **36**, 1850 (2003).

[50]  W. C. Oliver and G. M. Pharr, An improved technique for determining hardness and elastic modulus using load and displacement sensing indentation experiments, J. Mater. Res. **7**, 1564 (1992).

[51]  M. Fischer, M. Trant, K. Thorwarth, J. Patscheider, and H. J. Hug, A setup for arc-free reactive DC sputter deposition of Al-O-N, Surf. Coatings Technol. **362**, 220 (2019).

[52]  J. Malcolm W. Chase, *NIST-JANAF Thermochemical Tables* (Fourth edition. Washington, DC: American Chemical Society; New York: American Institute of Physics for the National Institute of Standards and Technology, 1998., 1998).

[53]  H. Schulz and K. H. Thiemann, Crystal structure refinement of AlN and GaN, Solid State Commun. **23**, 815 (1977).

[54]  P. Yang, H.-K. Fun, I. A. Rahman, and M. I. Saleh, Two phase refinements of the structures of α-$Si_3N_4$ and β-$Si_3N_4$ made from rice husk by Rietveld analysis, Ceram. Int. **21**, 137 (1995).

[55]  N. Ishizawa, T. Miyata, I. Minato, F. Marumo, and S. Iwai, A structural investigation of α-$Al_2O_3$ at 2170 K, Acta Crystallogr. Sect. B Struct. Crystallogr. Cryst. Chem. **36**, 228 (1980).

[56]  H. Banno, T. Hanai, T. Asaka, K. Kimoto, and K. Fukuda, Electron density distribution and disordered crystal structure of 15R-SiAlON, $SiAl_4O_2N_4$, J. Solid State Chem. **211**, 124 (2014).

[57]  T. Ban and K. Okada, Structure Refinement of Mullite by the Rietveld Method and a New Method for Estimation of Chemical Composition, J. Am. Ceram. Soc. **75**, 227 (1992).

[58]  J. M. Nieuwenhuizen and H. B. Haanstra, Microfractography of thin films, Philips Tech. Rev. **27**, 87 (1966).

[59]  R. N. Tait, T. Smy, and M. J. Brett, Modelling and characterization of columnar growth in evaporated films, Thin Solid Films **226**, 196 (1993).

[60]  C. S. Sandu, F. Medjani, R. Sanjinés, A. Karimi, and F. Lévy, Structure, morphology and electrical properties of sputtered Zr–Si–N thin films: From solid solution to nanocomposite, Surf. Coatings Technol. **201**, 4219 (2006).

[61]  M. Benkahoul, C. S. Sandu, N. Tabet, M. Parlinska-Wojtan, A. Karimi, and F. Lévy, Effect of Si incorporation on the properties of niobium nitride films deposited by DC reactive magnetron sputtering, Surf. Coatings Technol. **188**–**189**, 435 (2004).

[62]  S. Veprěk, The search for novel, superhard materials, J. Vac. Sci. Technol. A Vacuum, Surfaces, Film. **17**, 2401 (1999).

[63]  J. Musil, Hard nanocomposite coatings: Thermal stability, oxidation resistance and toughness, Surf. Coatings Technol. **207**, 50 (2012).

[64]  S. Veprek, Recent search for new superhard materials: Go nano!, J. Vac. Sci. Technol. A Vacuum, Surfaces, Film. **31**, 50822 (2013).

[65]  J. Procházka, P. Karvánková, M. G. J. Veprěk-Heijman, and S. Veprěk, Conditions required for achieving superhardness of ≥45GPa in nc-TiN/a-$Si_3N_4$ nanocomposites, Mater. Sci. Eng. A **384**, 102 (2004).

[66]  S. Veprek, P. Karvankova, and M. G. J. Veprek-Heijman, Possible role of oxygen impurities in degradation of nc-TiN∕a-Si3N4 nanocomposites, J. Vac. Sci. Technol. B Microelectron. Nanom. Struct. Process. Meas. Phenom. **23**, L17 (2005).

[67]  S. Veprek and M. G. J. Veprek-Heijman, Limits to the preparation of superhard nanocomposites: Impurities, deposition and annealing temperature, Thin Solid Films **522**, 274 (2012).

[68]  S. H. Sheng, R. F. Zhang, and S. Veprěk, Decomposition mechanism of $Al_{1-x}Si_xN_y$ solid solution and possible mechanism of the formation of covalent nanocrystalline AlN/$Si_3N_4$ nanocomposites, Acta Mater. **61**, 4226 (2013).

[69]  G. Liu, Z. Zhou, Q. Wei, F. Fei, H. Yang, and Q. Liu, Preparation and tunable optical properties of ion beam sputtered SiAlON thin films, Vacuum **101**, 1 (2014).

[70]   F. Anjum, D. M. Fryauf, J. Gold, R. Ahmad, R. D. Cormia, and N. P. Kobayashi, Study of optical and structural properties of sputtered aluminum nitride films with controlled oxygen content to fabricate Distributed Bragg Reflectors for ultraviolet A, Opt. Mater. (Amst). **98**, 109405 (2019).

[71]   R. D. Shannon and R. X. Fischer, Empirical electronic polarizabilities of ions for the prediction and interpretation of refractive indices: Oxides and oxysalts, Am. Mineral. **101**, 2288 (2016).

[72]   A. Bikowski, A. Holder, H. Peng, S. Siol, A. Norman, S. Lany, and A. Zakutayev, Synthesis and Characterization of (Sn,Zn)O Alloys, Chem. Mater. **28**, 7765 (2016).

[73]   J. Musil, R. Jílek, M. Meissner, T. Tölg, and R. Čerstvý, Two-phase single layer Al-O-N nanocomposite films with enhanced resistance to cracking, Surf. Coatings Technol. **206**, 4230 (2012).

[74]   T. F. Scientific, *Table of Elements: X-Ray Photoelectron Spectroscopy of Atomic Elements*, https://www.thermofisher.com/no/en/home/materials-science/learning-center/periodic-table.html.

# Supplementary information for

# Accelerating the development of oxynitride thin films: A combinatorial investigation of the Al-Si-O-N system

Stefanie Frick[1], Oleksandr Pshyk[1], Arnold Müller[2], Alexander Wieczorek[1], Kerstin Thorwarth[1], Sebastian Siol[1*]

[1] Empa, Swiss Federal Laboratories for Materials Science and Technology, Dübendorf, Switzerland

[2] Laboratory of Ion Beam Physics, ETH Zurich, Zürich, Switzerland

*Corresponding author:*
*Sebastian Siol, Sebastian.Siol@empa.ch*

## Materials and methods

**UHV transfer between deposition chamber and XPS system**

Due to the observation of an oxygen enriched surface layer for ex-situ samples in XPS, four libraries with low oxygen contents were transferred via a UHV transfer cart between the deposition chamber and the XPS system. During this procedure, the pressure never exceeded $7 \cdot 10^{-8}$ mbar. Next to a composition analysis of the surface for all 45 sample points per library, depth profiles were performed on eight sample points. A comparison between surface composition and after 10 and 15 min 2 kV $Ar^+$ sputtering is displayed in **Fig. S1**. The cation composition exhibits a linear correlation between measurements before and after sputtering as well as only a small surface enrichment of 2-4% Si/(Al+Si). Contrarily, the surface oxygen content exceeded the one after sputtering significantly with 8-12% O/(N+O). Moreover, a variation of the deposition temperature influenced the relation between the anion composition on the surface and after sputtering. Therefore, different trend lines were obtained for samples deposited at 400 °C and without heating. To ensure comparability among all sample libraries, the trend lines indicated in **Fig. S1** a) and b) were used to calculate cation and anion composition after sputtering.

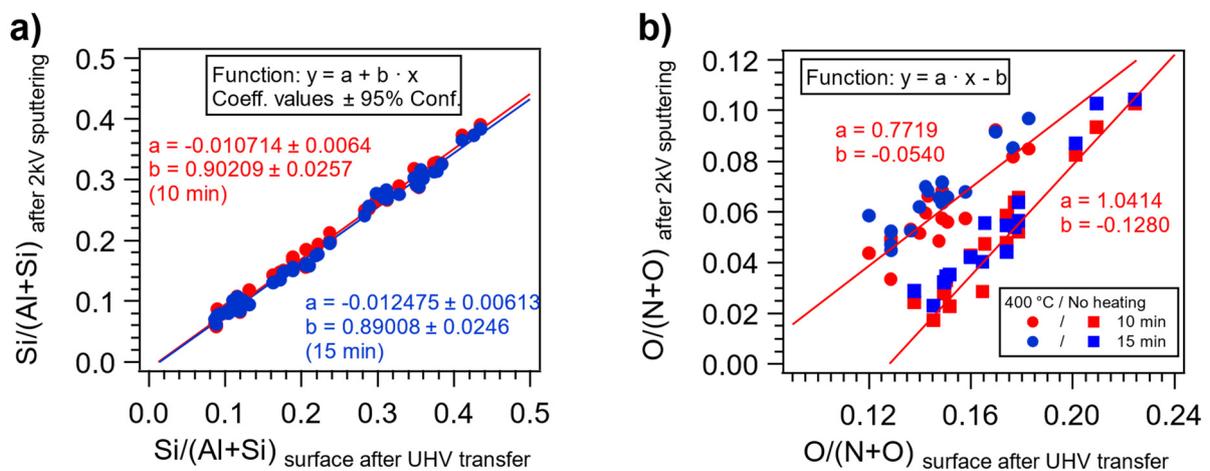

**Fig. S1.** Cation (a) and anion (b) composition after sputtering against those of the surface after UHV transfer with indicated trend lines.

**Calibration curve for anion composition**

XPS sputter depth profiles on both UHV transferred and ex-situ samples revealed an oxygen-rich surface layer. Therefore, the compositions reported throughout this study are based on XPS measurements performed after 10 min 2 kV $Ar^+$ sputtering. Since the sputtering procedure leads to the development of a low binding energy component in the Si 2p spectra, a slight reduction and as well as a change in the oxygen to nitrogen ratio cannot be excluded. Therefore, HI-ERDA measurements on selected samples were performed to calibrate the oxygen-to-nitrogen ratio obtained by XPS measurements after sputtering. The calibration curve and an exemplary resulting concentration-depth profile obtained from an ERDA measurement are displayed in **Fig. S2**. The ERDA profiles confirm the excess of oxygen at the surface as determined by XPS. With increasing film depth, the ERDA profile shows a homogenous elemental distribution. Therefore, ERDA compositions were determined by averaging over a depth range as indicated in **Fig. S2b** to exclude the effect of surface oxidation. The relative error of the presented ERDA measurement is of the order of 7% and 10% for H.

Assuming the uncertainty of the fitting parameters of the calibration curve presented in **Fig. S2a**, uncertainties between ~1 an.% (low oxygen contents) ~2 an.% (higher oxygen contents) are estimated applying Gaussian error propagation.

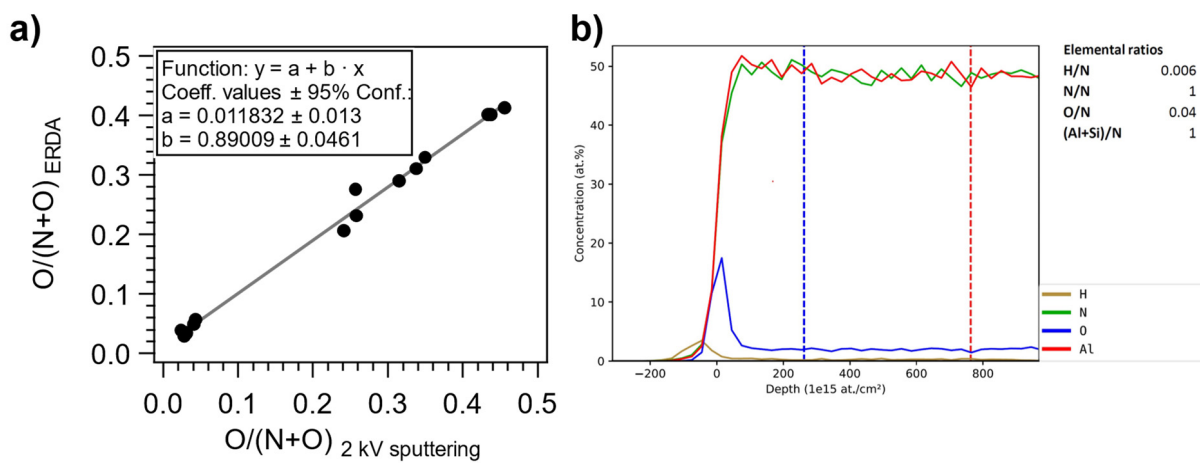

**Fig. S2.** a) Calibration of the O/(N+O) ratio between results from XPS after 10 min 2 kV sputtering and ERDA. b) Exemplary concentration depth profile from ERDA. An oxygen enriched surface layer is apparent.

**Determination of the optical constants and the thickness with the Swanepoel method**

As discussed by Poelman and Smet, the usage of only transmission spectra for the determination of the refractive index and the extinction coefficient has several advantages, e.g. no requirement for accurate reflection data and no requirement for dispersion model assumptions and sensitivity to surface contamination layers as in the case of ellipsometry [49]. Therefore, the envelope method developed by Manifacier *et al.* [48] and Swanepoel [47] was implemented in a MATLAB® code. The method can be applied for single layer films on a non-absorbing substrate in a range of low absorption (T ≥ 70%). As visualized in the main text Fig. 4, next to the transmission spectrum of the system film on substrate the transmission spectrum of the substrate or its wavelength dependent refractive index is required (both options are implemented in the script). From the interference maxima and minima, envelopes are constructed, which allow for the determination of the refractive index, the absorption/ extinctions coefficients and the thickness of the sample. In the implemented code, the envelope is constructed via a built-in interpolation function employing the modified Akima interpolation, leading for our spectra to less undulations for thinner samples with few, largely spaced fringes in comparison to simple cubic or spline interpolation. Since the envelope should ideally be tangential to the transmission curve [47,49], the error of the determined envelope in regions with pronounced drops in transmission will be significantly higher in the implemented version of the script. To avoid the incorrect detection of the extrema due to outliers, noisy spectra can be smooth a built-in function. Applying smoothing, e.g. with a Savitzky-Golay filter, care is advised by choosing parameters, like frame size, to preserve the fingers height and position. Moreover, Manifacier *et al.* and Swanepoel derive their models slightly differently assuming an infinite respectively a finite substrate. In the limit of low absorption, this difference influences mainly the determined absorption coefficients [47]. Both variants are implemented in the script and can be selected. Moreover, Swanepoel suggested further corrections to obtain higher accuracy for thickness and refractive index by determining the interference order number of the extremum. The correction can be selected in the script as well, while the determination of the order number was determined via the "paper and pencil" calculation suggested by Poelman and Smet [49]. For the determination of refractive indices presented in this paper, the latter correction was not employed, only for the determination of the thickness. Moreover, the finite substrate assumption was applied, as well as a Savitzky-Golay filter with order 2 and frame length 31 (~8 nm). This led to decent smoothing for the region > 500 nm, while transmission extrema in the region around 400 nm were slightly flattened, which reduces the refractive index in this region possibly by up to 0.005 and increased the absorption coefficient up to ~1%. Based on the noise in the smoothed transmission curves, an uncertainty of the refractive index at 633 nm of 0.015 is estimated for samples with thicknesses > 800 nm.

**Nanoindentation: Maximal indentation depth**

In order to decide on a maximum indentation depth, a procedure based on the ISO standard (ISO 14577-4:2016) was chosen. Six multicycle indents at 10 different depths on each corner of the library were performed to test maximal and minimal thicknesses. Examples of typical multicycle results (maximal load of 50 mN) are shown in **Fig. S3**. Independent on the thickness, a significant spread of the hardness, with a tendency to increased hardness values with decreasing depth, were found. For the sample with a thickness of ~900 nm, a decay in hardness occurs above 200 nm contact depth presumably associated with deformation in the substrate, while no decay was observed for samples with thickness of ~1550 nm. To ensure neither seeing the influence of the substrate nor overestimating the hardness due to a low indentation depth, the maximal indentation depth was chosen to 150 nm (contact depth ~130 nm) to fall in an approximately constant hardness region.

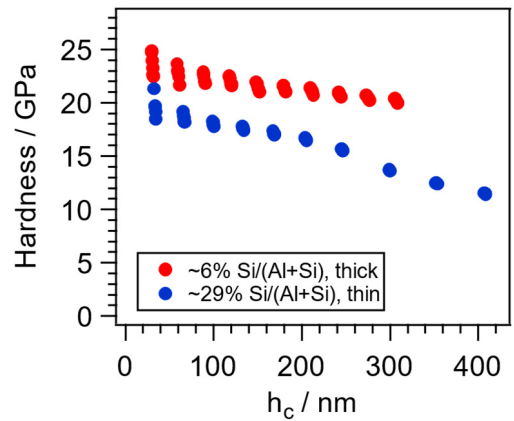

**Fig. S3.** Depth dependent hardness for a thick (~1550 nm) and a thin (~900 nm) sample.

**30 V RF substrate bias during deposition: Ar distribution map**

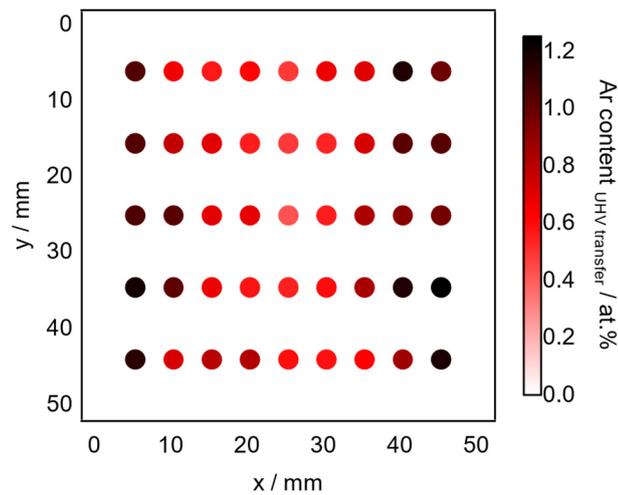

**Fig. S4.** Argon content in a combinatorial library deposited with applied RF substrate bias of 30 V. The values are obtained by XPS measurements of the surface after a UHV transfer of the library between deposition chamber and XPS. The left and right edge of the library face towards the metallic Al and Si target.

# Results and discussion

## Approaches for O/N-gradients

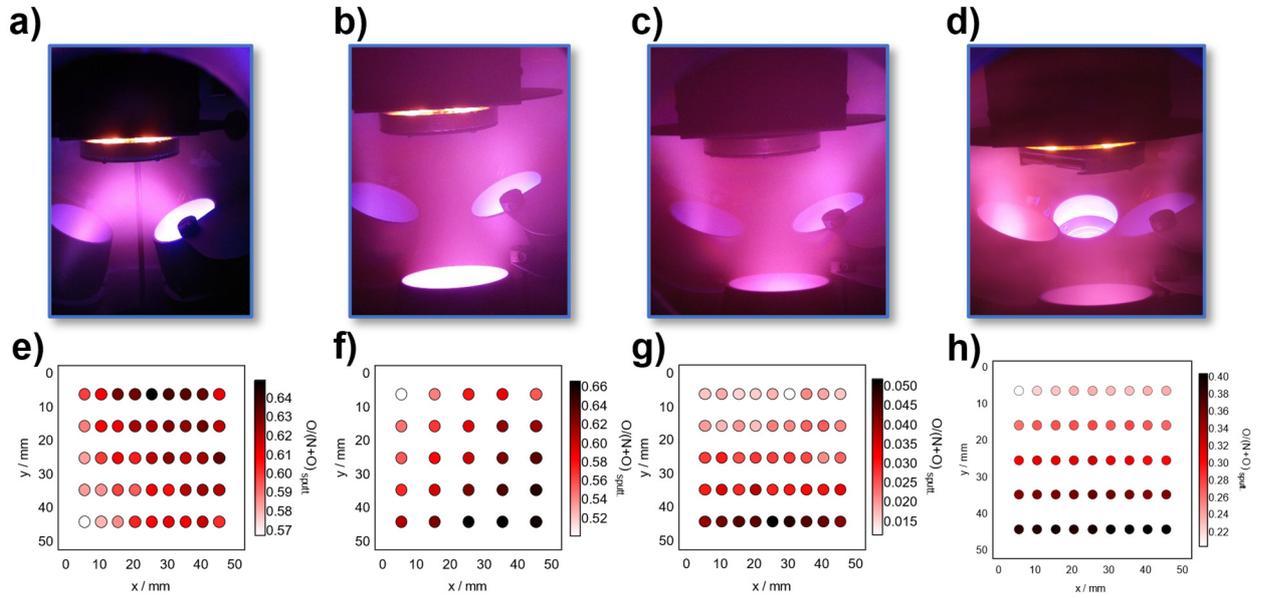

**Fig. S5.** a)-d) Photographs of the four different setups to obtain O/N gradients as presented in the main text Fig.1. e)-h) Exemplary O/N distribution maps obtained by the different approaches.

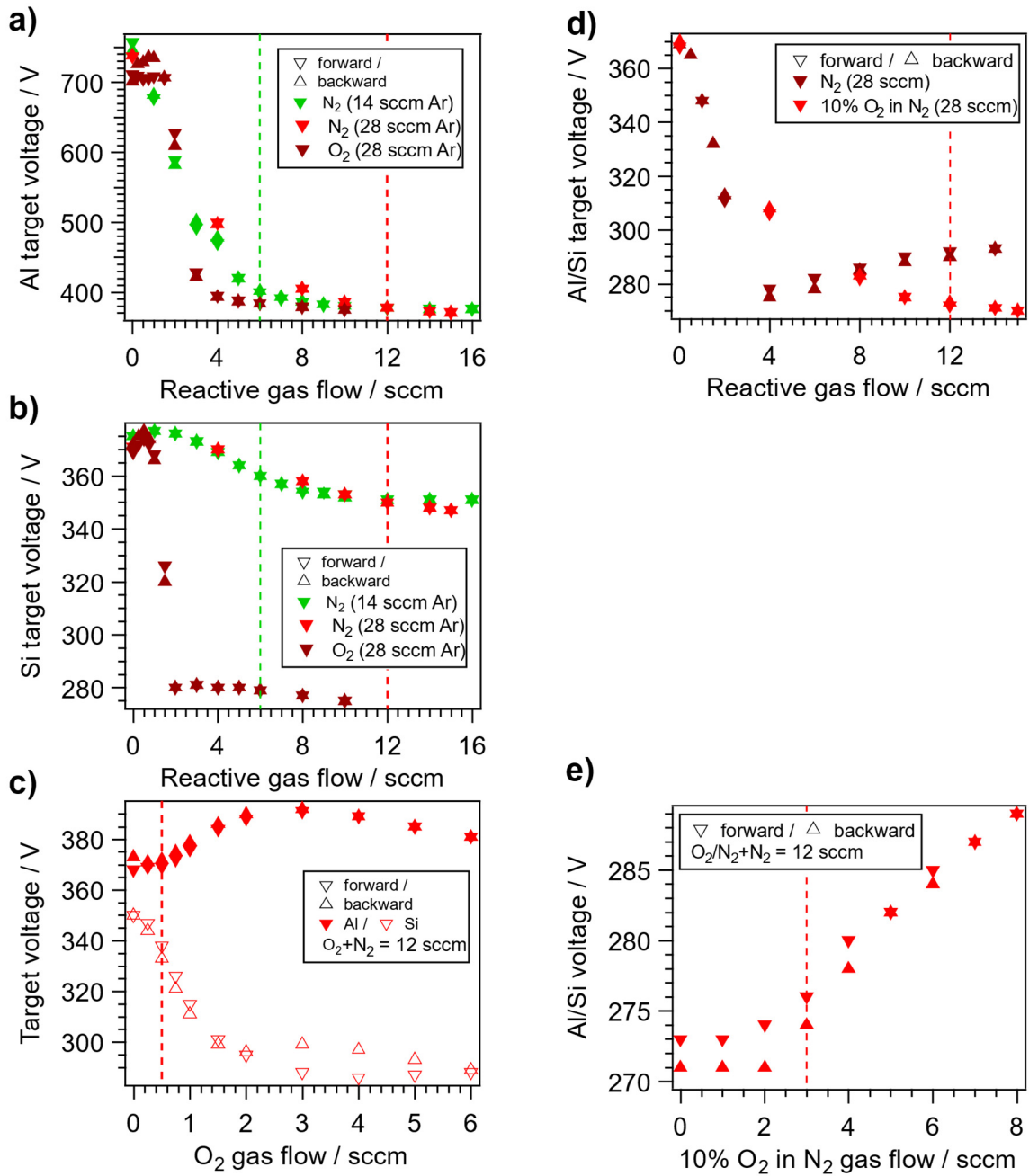

**Fig. S6.** Hysteresis experiments: a), b) and c) for the first setup (with $O_2$ supply via a nozzle) and d) and e) for the second setup (with additional Al/Si 80/20 target). The total pressure and Ar flow were kept constant as for the respective depositions. a), b) and d) show hysteresis curves of only one reactive gas per experiment (either $N_2$, $O_2$ or a mix gas). The curves in c) and e) were recorded at a constant total reactive gas flow of 12 sccm, while the $O_2$ or mix gas content was varied. Dashed red lines represent the respective flows employed for the presented materials libraries in **Fig. 1e,f**. The green curves in a) and b) are recorded for Ar flows set for the third approach (oxygen gradient by sputtering from oxide targets). Green dashed lines represent the $N_2$ flow applied for **Fig. 1g,h** and all libraries for the property screening of the Al-Si-O-N system.

**Tab. S1.** Standard formation enthalpies of aluminum and silicon oxides and nitrides from the NIST-JANAF database [52].

|  | AlN | $Al_2O_3$ | $Si_3N_4$ | $SiO_2$ |
|---|---|---|---|---|
| $\Delta H_f°$ / kJ/mol per cation | -317.984 | -837.846 (α) | -248.251 (α) | -910.857 (Quartz) |

**Composition screening**

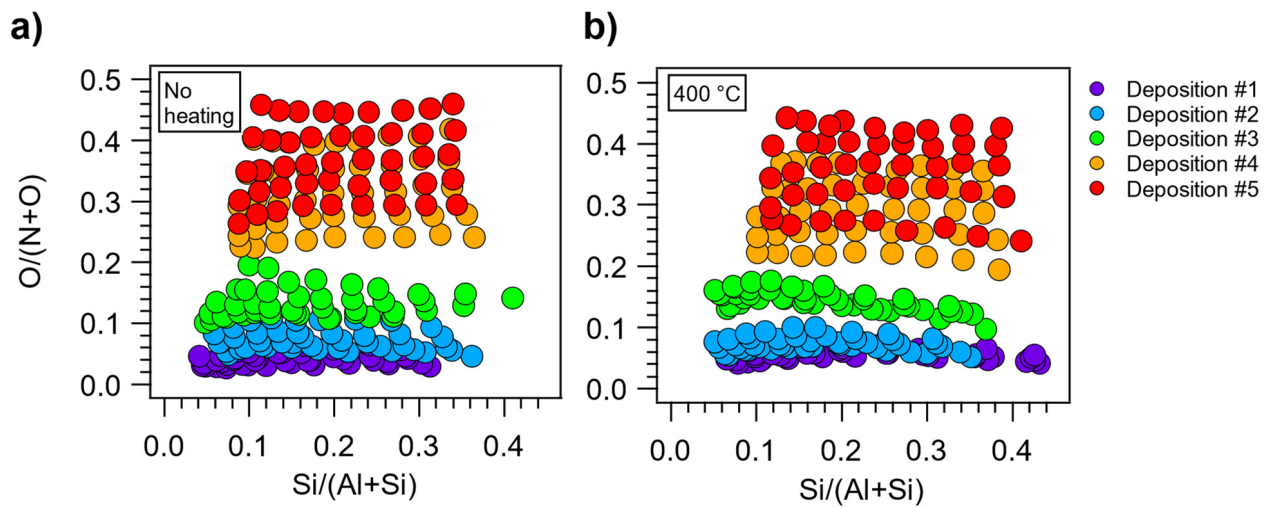

**Fig. S7.** Composition spreads obtained with different compositions deposited without external substrate heating (a) and at 400 °C (b). Depositions #1 to #3 were deposited using only an $Al_2O_3$ target as oxygen source (see **Fig. 1c**), while for depositions #4 and #5, both an $Al_2O_3$ and a $SiO_2$ target were employed (see **Fig. 1d**).

## XRD lattice parameter and Ψ rocking curves

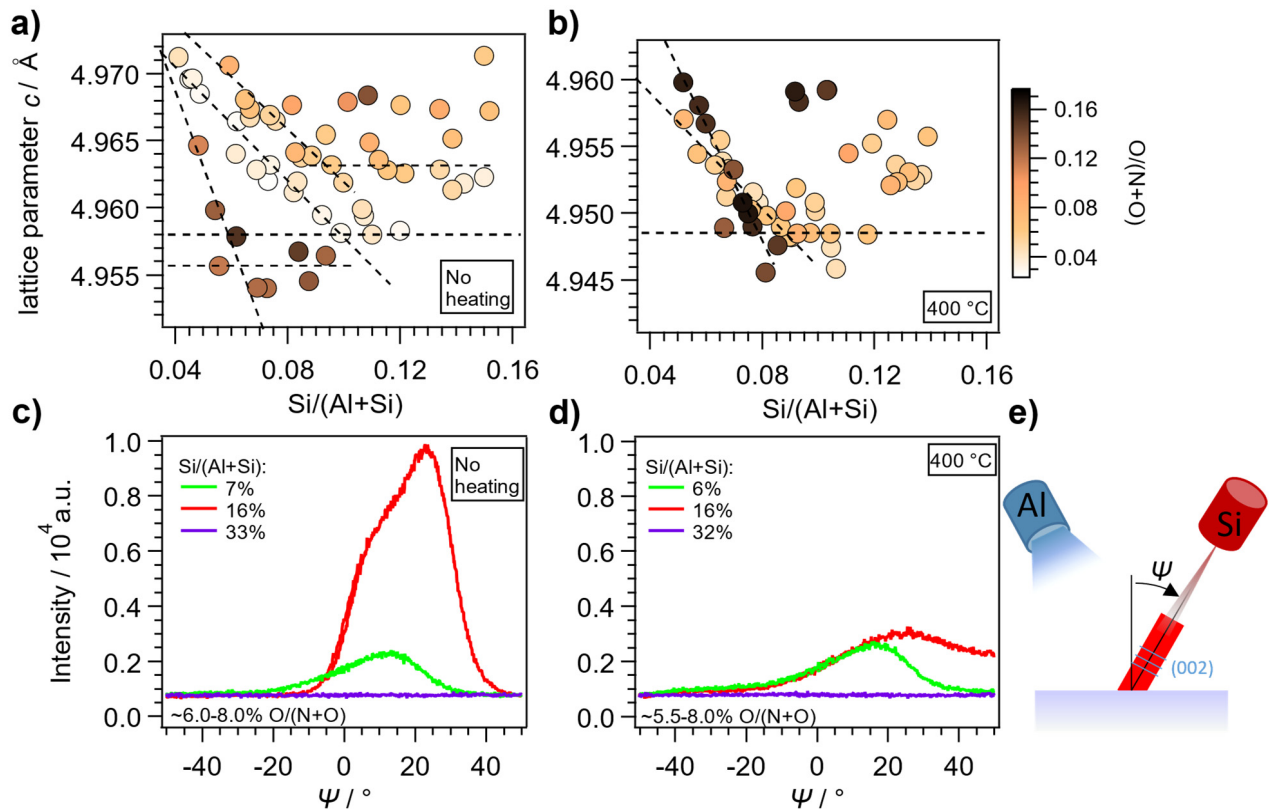

**Fig. S8.** Lattice parameter c as calculated from the (002) peak position (a),b)) and Ψ rocking curves of three points with varying silicon contents and similar oxygen contents to assess the preferential orientation of the (002) lattice planes (c), d)). The (002) reflection angle was fixed to the peak maximum observed in Θ/2Θ scans. The geometrical meaning of the angle Ψ with respect to the sputter setup is indicated in e).

The lattice parameter $c$ estimated from the (002) w-AlN reflection position indicates a solubility limit of approximately 8-9 cat.% silicon 400 °C, situated inside the w-AlN region (c.f. **Fig. 2**). For only few cation percent Si, a constant lattice parameter is obtained, before the development of the low angle shoulder leads to significant peak shifts. The range reduces even with increasing oxygen content. This could indicate a very limited extension of the nanocomposite region with a monolayer thick grain boundary phase.

**Nitrogen polarizability**

The polarizability of the nitrogen ions was estimated by applying the Anderson-Eggleton equation as given by Shannon and Fischer to calculate the total dynamic polarizability of the compounds [71]:

$$\alpha_{\mathrm{AE}} = \frac{V_{\mathrm{M}}}{\frac{4\pi}{n^2 - 1} + \frac{4\pi}{3} - 2.26}$$

$\alpha_{\mathrm{AE}}$ represents the dynamic polarizability of the compound, $V_{\mathrm{M}}$ the molar volume and $n$ the refractive index. The number 2.26 accounts for a certain electron overlap [71], i.e. a certain degree of covalency. Furthermore, the compound polarizability can be approximated by a linear combination of electronic polarizabilities $\alpha_{\mathrm{e},i}$ of the different species $i$ weighted by their number per unit cell $m_i$ [71]:

$$\alpha_{\mathrm{AE}} = \sum_i m_i \alpha_{\mathrm{e},i}$$

We assumed a unit cell volume of AlN and calculated the oxygen polarizabilities as a function of the Si/(Al+Si) ratio (linear function between the polarizabilities of $SiO_2$ and $Al_2O_3$). The specific parameters are given in **Tab. S2**. In **Fig. S9**, the resulting nitrogen polarizabilities are displayed as a function of the cation and anion composition.

**Tab. S2.** Parameters employed to estimate the nitrogen polarizability from the refractive index and the composition by applying the Anderson-Eggleton equation. The polarizability values were determined for a wavelength of 589.3 nm.

| $V_{\mathrm{M}}$ / Å³/atom | $\alpha_{\mathrm{Al}^{3+}_{[4]}}$ / Å³ | $\alpha_{\mathrm{Si}^{4+}_{[4]}}$ / Å³ | $\alpha_{\mathrm{O},Al_2O_3}$ / Å³ | $\alpha_{\mathrm{O},SiO_2}$ / Å³ |
|---|---|---|---|---|
| 10.4 [53] | 0.533 [71] | 0.284 [71] | 1.51 [71] | 1.58 [71] |

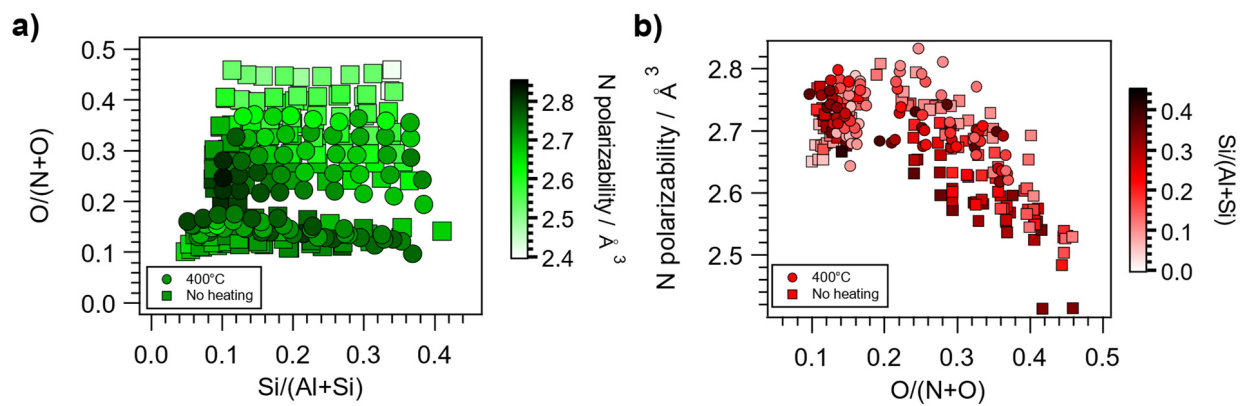

**Fig. S9.** Nitrogen polarizability as a function of cation and anion composition.

**Absorption coefficients and visual appearance**

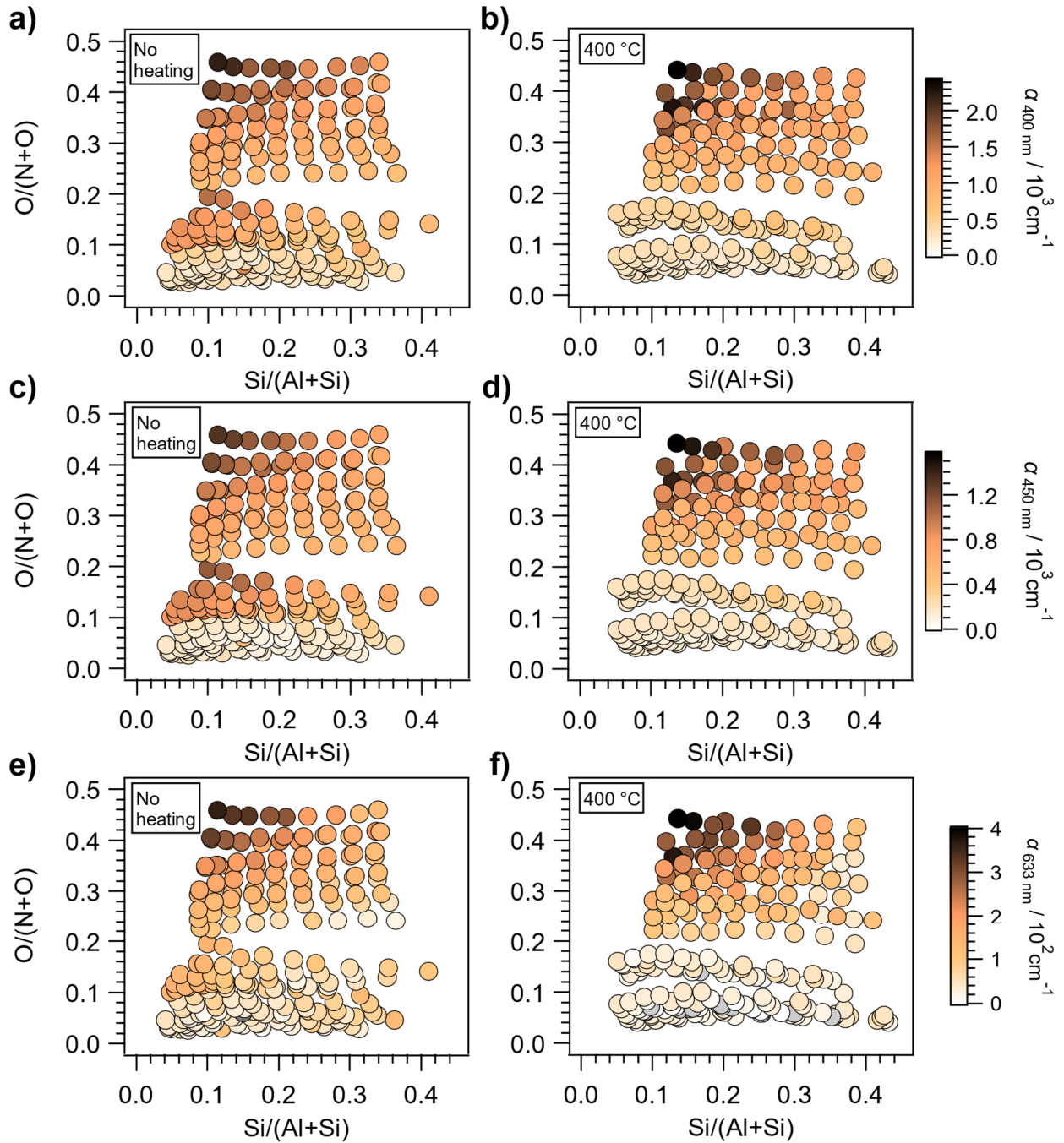

**Fig. S10.** Absorption coefficients obtained by Swanepoel's envelope method at different wavelengths: 400 nm (top), 450 nm (middle), 633 nm (bottom). Grey points indicate negative values, which can occur within the error of the measurement and analysis, if the samples are highly transparent.

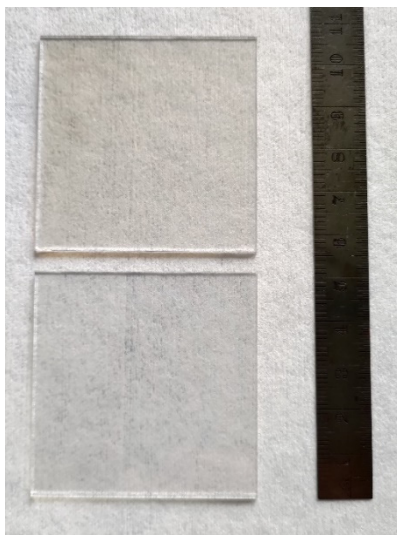

**Fig. S11.** Photograph of Al-Si-O-N libraries on glass substrates deposited at 400 °C with 26-46% O/(N+O) and 9-34% Si/(Al+Si) (top) and 3-6% O/(N+O) and 4-31% Si/(Al+Si) (bottom). Library samples on the left are Al-rich, on the right Si-rich.

**XPS core level peak fitting & anion-cation binding energy differences**

Peak fitting of the core level spectra was performed in the CasaXPS software. A Shirley background was applied, for the Si 2p spectra with an additional offset. Gaussian-Lorentzian peak shapes were chosen for the O 1s (GL(40)), N 1s (GL(50)), Si 2p (GL(30)) and Al 2p (GL(40)) to extract binding energy differences between the different elements. All spectra were fitted with one component, except for the Si 2p spectra, where a low binding energy component occurs presumably due to reduction during sputtering. Doublet splitting of 0.44 eV and 0.65 eV were applied for the Al $2p_{3/2}$-Al $2p_{1/2}$ and the Si $2p_{3/2}$-Si $2p_{1/2}$ spectra were applied based on the Thermo Fisher Database [74]. Representative spectra are depicted in **Fig. S12**.

XPS core level positions are a useful tool to analyze the chemical state of elements: The binding energy (BE) of the core level of an element typically increases with increasing oxidation state. To ensure insensitivity to charge referencing issues, binding energy differences between cation and anion core levels were investigated as a measure for the bond covalency/ionicity. N 1s and Al 2p core levels were chosen as they are the main constituents of the compounds, allowing for more reliable fitting results. With increasing covalency of a bond, valence electrons are shared to a higher extent. Consequently, the electron density close to the cation is increased (binding energy decreases), while the electron density reduces at the anion (binding energy rises). Following this concept, the binding energy difference between N 1s and Al 2p should rise in case of a higher covalency. In Fig. S12**Fig. S13**, the functional properties are plotted against this very binding energy difference. Both refractive index and effective Young's modulus, as well as the hardness in the X-ray amorphous region exhibit a tendency to increase with the binding energy difference. Instead of a narrow linear correlation, a rather broad band is observed in all three cases. This deviation is based on different dependencies on the silicon cation content: While the binding energy difference increases with higher silicon content (as expected from increasing covalency due to silicon's higher electronegativity in comparison to aluminum), the three functional properties decrease. Further research will be required in order to understand what additional materials property causes this (apparent) correlation among properties besides covalency, e.g. variations in material density.

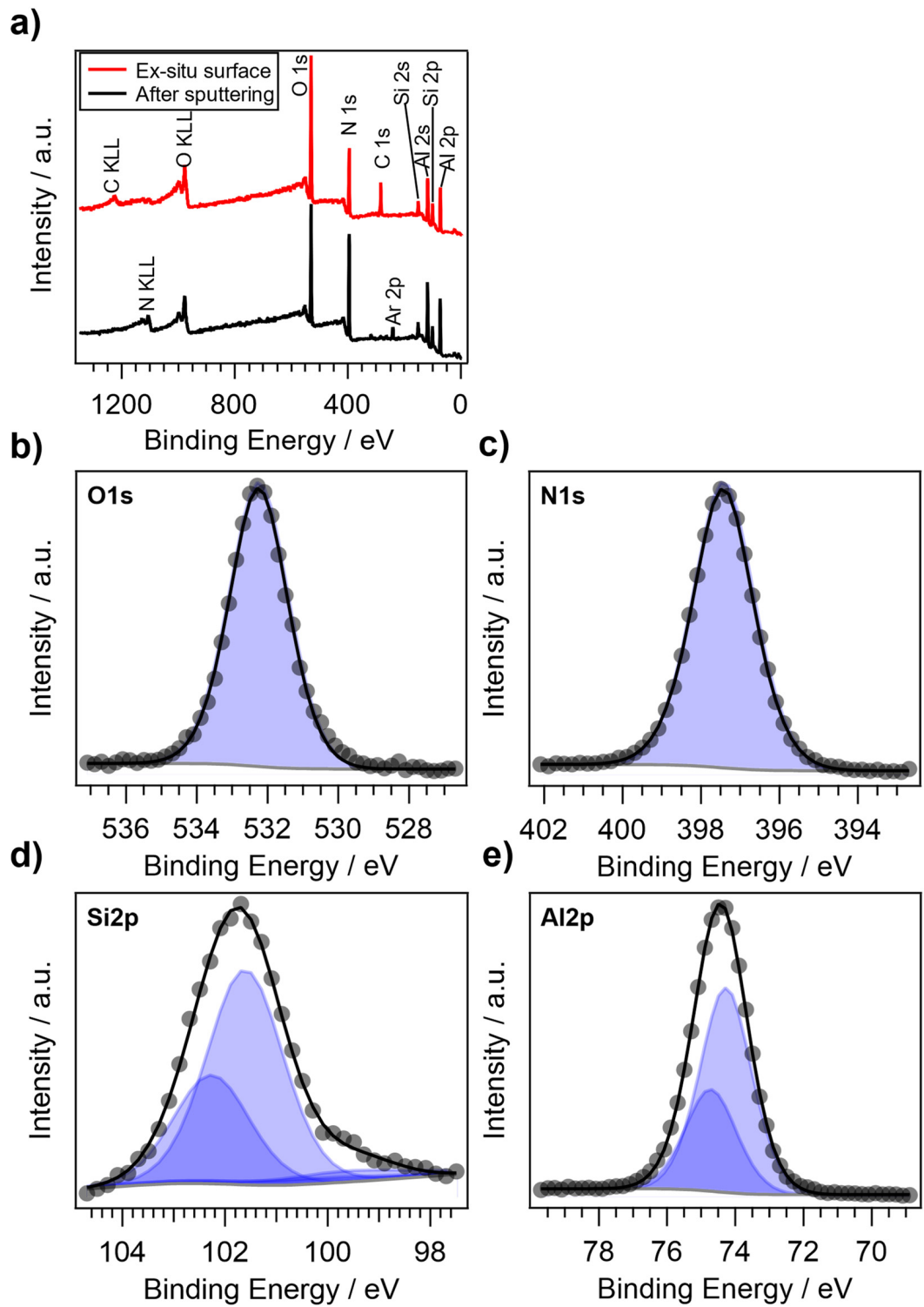

**Fig. S12.** Exemplary survey spectra before and after 10 min 2 kV sputtering (a) and core level spectra after sputtering with respective Gaussian-Lorentzian peak fits (b-f).

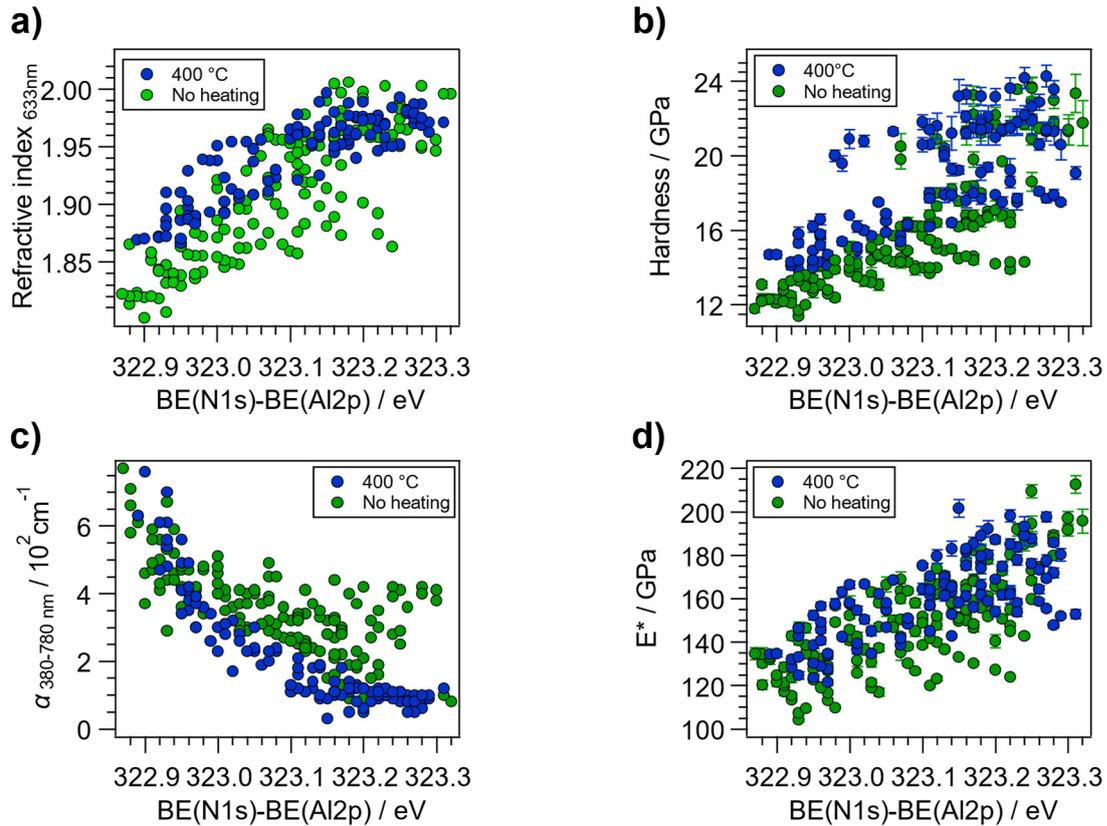

**Fig. S13.** Correlation of the XPS binding energy difference between N 1s and Al 2p peaks after sputtering to a) the refractive index at 633 nm, b) the hardness, c) the mean absorption coefficient in the visible range and d) the effective Young's modulus.